\documentclass[11pt,a4paper]{article}

\usepackage{amsmath,amsthm,amssymb}

\usepackage{euscript}

\usepackage{graphics,graphicx}
\usepackage{epsfig}
\usepackage{multicol}
\usepackage{color}
\makeatletter
\@addtoreset{equation}{section}
\makeatother

\begin{document}

\begin{center}
{\Large\bf Gauge field back-reaction in Born Infeld cosmologies}
\end{center}
\vspace{0.5cm}
\begin{center}
\large{\bf Paulo Vargas Moniz,} ${}^{a,}$\footnote{pmoniz@ubi.pt;
Also at CENTRA-IST, Lisboa, Portugal} \large{\bf and John Ward}
${}^{b,}$\footnote{jwa@uvic.ca}
\end{center}
\begin{center}
\emph{ ${}^a$ Departamento de F\'{\i}sica, Faculdade de Ci\^encias,
UBI, 6200 Covilh\~{a}, Portugal
\\} \vspace{0.1cm} \emph{${}^{b}$ Department
of Physics and Astronomy, University of Victoria, Victoria, BC \\
V8P 1A1, Canada }
\end{center}
\begin{abstract}
In this paper, we investigate the back-reaction of $U(1)$ gauge
fields into a specific class of
inflationary settings. To be more precise, we employ a Bianchi-I
geometry (taken as an anisotropic perturbation of a flat FRW model)
within two types of Born-Infeld theories. Firstly, we consider pure
Born-Infeld electromagnetism. For either a constant or a $b(\phi)$
coupling, inflationary trajectories are modified but anisotropies
increase; In particular, for the former coupling we find that a
quadratic inflaton potential, within a constant ratio for the scalar
and gauge energy densities, does not induce sufficient inflation,
while in the latter the back-reaction in the cosmology determines
(from the tensor-scalar ratio) a narrow range where inflation can
occur. A Dirac-Born-Infeld framework is afterwards analysed in both
non-relativistic and relativistic regimes. In the former, for
different cases of the coupling (richer with respect to mere BI
setups) between scalar and gauge sectors, we find that inflationary
trajectories are modified, with anisotropy increasing or decreasing.
In particular, a tachyonic solution is studied, allowing for a non
standard ratio between scalar and gauge matter densities, enhancing
sufficient inflation, but with the anisotropy increasing. For the
relativistic limit, inflationary trajectories are also modified and
anisotropies increase faster than in the non-relativistic limit.
Finally we discuss how magnetic seed fields could evolve in these
settings.
\end{abstract}
\newpage
\section{Introduction}
\label{Intro}

\indent Modern cosmology has entered a new phase of precision,
unthinkable only two decades ago \cite{Primack:2006it}. However a
fundamental issue remains open; namely the origin of galactic and
cosmological magnetic fields. Such fields have coherence length of
the order $1$ Mpc, with a magnitude of around $10^{-7} G$
\cite{Grasso:2000wj, Giovannini:2003yn}. The current explanation for
the origin of such fields is that a small seed field emerged during
primordial inflation, which was amplified at late times by the
galactic dynamo mechanism \cite{Turner:1987bw, Enqvist:1998fw}; Cf.
\cite{Tsagas:2009cr, Matarrese:2004kq,
Tsagas:2001ak,Kandus:2010nw,Kunze:2009bs} and references therein for
additional elements on cosmic magnetogenesis. As is well known from
cosmological perturbation theory, the accelerated expansion
stretches small scale physics to cosmological size, therefore it
would appear natural to include a primordial magnetic field (e.g., a
fluctuation) in models of early universe cosmology. A crucial
feature in this setting is that a FRW cosmology with canonical
fields is conformally invariant, therefore a mechanism to break the
conformal invariance is necessary for magnetic fields to be present
\cite{Dolgov:1993vg}. Proposals of solutions to this problem are
discussed in \cite{Giovannini:2003yn, Enqvist:1998fw}.


Additionally, dealing with a vector (gauge) field in cosmology may
require some degree of anisotropy\footnote{See ref.
\cite{MMS,Dyadichev:2001su} for Einstein-Yang-Mills configurations
consisting of FRW settings with non-Abelian (e.g., $SU(2)$) gauge
fields.} \cite{Bennett:2010jb}. Therefore we must go beyond the
simplest FRW models to understand the dynamics of the universe with
a magnetic field. In this context, let us mention the various
anomalies present in the CMB, such as the suppression and
asymmetries of the power spectrum in different hemispheres, together
with  the presence of a preferred direction due to the alignment of
lowest multipole moments \cite{Land:2005ad, Copi:2006tu}. These
anomalies can be interpreted as a sign that isotropic statistics are
only a leading order approximation, and that these anomalous
measurements are due to non-isotropic expansion. Hence, it is
reasonable to investigate (inflationary) anisotropic space-times
with a non-zero gauge field.

Model building with both magnetic seed fields and anisotropies has
been explored in \cite{Kahniashvili:2008hx, Kanno:2009ei,
Gumrukcuoglu:2010yc, Emami:2009vd, Dulaney:2010sq}. One potential
pitfall present in those references was that the magnetic field
energy density grows rapidly during an inflationary phase, and was
thought to destroy the inflationary dynamics. As pointed out in
\cite{Kanno:2009ei}, this claim was not based on concrete analysis
and in fact was shown to be false. Inflation does occur in the
presence of a gauge field and leads to weakly growing anisotropy -
but is modified by the back-reaction on the inflaton
\cite{Watanabe:2009ct,Watanabe:2010fh}. This was due to the gauge
field acting as a source term in the Klein-Gordon equation, whilst
remaining subdominant to the scalar energy density in the Hubble
parameter. The overall net effect was to further suppress the
amplitude of magnetic field production after inflation. The above
analysis refers to (canonical) Einstein-scalar-Maxwell theory.
However within the context of open string theory, the effective
action is non-linear and we can ask how the above results are
modified by such non-linearities.

Our purpose in this paper is to extend the discussion on gauge
fields back-reacting on inflationary cosmologies (and how magnetic
seed  fields can emerge and evolve consequently)  to a class of
non-linear theories described by Born-Infeld (BI) type Lagrangians.
Within string theory, we have  a Lagrangian description of the low
energy regime of a (Dirichlet) $Dp$-brane, which in a
cosmological context gives rise to
a class of theories known as Dirac-Born-Infeld (DBI)
inflation\footnote{In the simplest of such models, our universe can
be localised on a $D3$-brane which inflates as it
evolves through a warped internal
geometry. If such a model is to be correct, then there must also be
a non-zero gauge field on the world-volume - corresponding to
excited (Fundamental) $F$-strings.}. The inclusion of gauge fields
in such models have been considered in \cite{Ganjali:2005sr,Bamba:2008my}, but here we  incorporate instead the ideas of
anisotropic inflation and attempt to generalise the results. Moreover these papers
assumed warped backgrounds which were asymptotically AdS, however more
general classes of backgrounds are not asymptotically AdS which leads to
non-trivial coupling between the gauge field and the dilaton. By assuming an extra
condition, which we believe to be physically motivated, fixing the electromagnetic
energy density in terms of the scalar energy density - we can consider more general
configurations.

Our  paper is presented as follows. In section 2 we
consider a simple BI theory, which reduces to
the Einstein-Maxwell theory in a
specific limit. In subsection 2.1 we take the usual BI configuration
and in subsection 2.2 we  analyse the gauge and scalar sectors
through a $b(\phi)$ coupling. In both cases, we  investigate how the
gauge field back-reaction affects the inflationary dynamics. In
particular, we describe how inflation can be modified and whether
the anisotropies will increase or not.
In section 3 we employ a
phenomenological model of DBI inflation using a generalised
expression for the $D3$-brane action. We consider the back-reaction
of the gauge field in several different regimes. In particular, the
energy density can be constrained so that inflation is driven by the
scalar sector.  In subsection 3.1 we discuss the non-relativistic
limit, together with a tachyon configuration followed by  a $AdS_5$
solution. In subsection 3.2 we take the relativistic limit. In
section 4 we conclude with a discussion of our results and
suggestions for future work. In Appendix A  we determine how
magnetic seed fields can be generated and how DBI ingredients can
contribute to the magneto cosmogenesis discussion.


\section{Born-Infeld theory}
\label{BI}

\indent

In this section  we consider the non-linear extension of the
Einstein-Maxwell theory ($U(1)$ gauge field) to the
Einstein-Born-Infeld theory. Such a theory automatically includes
electro-magnetic duality, and a causal structure that implies that
the electromagnetic field is everywhere finite. Other works
detailing non-linear Lagrangian densities include
\cite{MosqueraCuesta:2009tf, Vollick:2001zd, Kunze:2007ph,
Campanelli:2007cg}. A similar framework plays an important role in
the effective description of $D$-branes within string theory, and
therefore serves as a toy model for a more detailed analysis of open
string dynamics.

The matter Lagrangian for such a model can be written as follows
\begin{equation}
\mathcal{L}= -\frac{1}{2} \partial_{\mu} \phi \partial^{\mu}\phi -
V(\phi) + \frac{1}{b}\left(1-\sqrt{1+\frac{b}{2}F_{\sigma
\rho}F^{\sigma \rho}} \right),
\end{equation}
where $b$ is the BI parameter. One
can see that as $b \to 0$ the
theory is reduced so that it includes the Maxwell form; However the
non-linear nature of the Lagrangian means that we can think of $b$
as a deformation parameter which has mass dimension $M^{-4}$.
Cosmologically therefore, we expect $b$ to deform the conformal
nature of the gauge field term. We wish to study the implications of
such a gauge field correction on the inflationary dynamics.

Using the  residual gauge invariance, we consider solutions where
$A_0=0$ and assume that the homogeneous vector potential
$A$ is aligned (and fixed) along
the $x$-direction. This breaks the (spatial) $SO(3)$ symmetry of the
space-time to $U(1) \times SO(2)$, with the planar symmetry
preserved transverse to the gauge field direction. Since our gauge
field breaks isotropy, we must ensure that the background metric is
anisotropic for the theory to be consistent \cite{Copi:2006tu, Kanno:2009ei, Campanelli:2009tk}.
It is simplest to
extend the FRW metric to a Bianchi I metric of the form
\begin{equation}
ds^2 = -dt^2 + e^{2\alpha(t)}[e^{-4 \sigma(t)}dx^2 +
e^{2\sigma(t)}(dy^2 + dz^2)], \label{B-I}
\end{equation}
where $a(t) = e^{\alpha}$ is the scale factor of the homogeneous
universe and $\sigma$ is herein a (perturbative) deviation from
isotropy. Let us be more concrete
about this point: In most of the case studies investigated in this
paper, our starting point will be that of a
 spatially flat FRW space-time filled with a scalar
field and a \emph{weak} non-linear electromagnetic field;
The energy density of the latter will be sought to be initially well
below that of the scalar field matter to ensure that eventually the
subsequent  evolution  is that of a Bianchi-I, but very close (in a
perturbative sense) to a flat FRW.
The above metric can thus be interpreted as describing a
back-reacted geometry (sourced by the vector potential); When
considering an initial state of isotropic inflation, one can
consistently set $A, \sigma \to 0$, in which case the metric reduces
to that of flat FRW.

The field equations for the gauge
field derived from the Lagrangian can then be written
\begin{equation}
0= \partial_{\mu} \left(\sqrt{-g} F_{ab}g^{\nu a} g^{\mu
b}(1+\frac{b}{2}F_{\sigma \rho}F^{\sigma \rho})^{-1/2} \right),
\end{equation}
however with the above ansatz for the gauge field, one sees that
the Maxwell equation admits the following solution
\begin{equation}
\dot{A}_x^2 \simeq \frac{P_A^2 e^{- 2\alpha-8\sigma}}{(1+ b P_A^2
e^{-4\alpha-4\sigma})},
\end{equation}
where $P_A$ is an integration constant, which we
can associate  with the
electromagnetic density. Note that in the limit $ b \to 0$ the above
solution reduces to that of Einstein-Maxwell theory. Because of the
unique structure of the BI Lagrangian, there is a causal bound on
the gauge field which can be written as
\begin{equation}
b \dot{A_x}^2 \le e^{2\alpha-4\sigma}
\end{equation}
and if this bound is saturated, then the gauge field term completely drops out
of the action leaving a residual cosmological constant term set by $1/b$.
In the isotropic scenario, the bound allows for a larger
gauge field energy density at late times. Therefore the initial
conditions for inflation are clearly sensitive to the magnitude of the
gauge field.
Calculation of the Einstein and scalar field equations gives us the
following expressions, written in the on-shell formalism for the gauge field
\begin{eqnarray}
3 \dot{\alpha}^2 - 3\dot{\sigma}^2 &=&
\frac{1}{M_p^2}\left(\frac{1}{2}\dot{\phi}^2 + V - \frac{1}{b} +
\frac{\sqrt{1 + b P_A^2 e^{-4\alpha-4\sigma}}}{b} \right), \label{2.7}\\
\ddot{\sigma}+3\dot{\alpha}\dot{\sigma} &=& \frac{1}{6 M_p^2 b}
\left(2 - \frac{(2-bP_A^2 e^{-4\alpha-4\sigma})}{\sqrt{1+bP_A^2
e^{-4\alpha-4\sigma}}} \right), \label{2.8}\\
2\ddot{\alpha}+3\dot{\alpha}^2 + 3\dot{\sigma}^2 &=&
\frac{1}{M_p^2}\left(-\frac{1}{2}\dot{\phi}^2 + V +
\frac{1}{3b}\left(1-\sqrt{1+b P_A^2 e^{-4\alpha-4\sigma}} \right)
\right), \label{2.9}\\
\ddot{\phi} + 3\dot{\alpha}\dot{\phi} + V'
&=& 0 \label{2.10},
\end{eqnarray}
which form the basis of our analysis;
Note however, that we assume $b$ to be constant here. By relaxing this constraint
the inflaton field equation becomes modified. This will be explored in section 2.2.
In what follows, we will be interested in inflationary solutions of the
Einstein equations in the regime of slow-roll and small
anisotropies. 


\subsection{Standard BI configuration}
\label{2.1}

\indent

Let us  consider  isotropic
inflation as our initial setting. We therefore
take the gauge field and $\sigma$
to zero, and the metric reduces to a flat FRW-form. Once the scalar
potential is specified, the resulting dynamics can be solved for
exactly. We will consider a small class of (chaotic) inflationary
potentials for simplicity, such that $V \sim m^p \phi^p /p$ where
$p$ is unspecified.
The scalar field equation can be solved in the
Hamilton-Jacobi formalism to obtain
\begin{equation}
\phi^2 (\alpha) = \phi_0^2 - 2 p \alpha M_p^2.
\end{equation}
Concretely for $p=2$ we would then find that inflationary solutions
using the WMAP normalisation, correspond to $m \sim 10^{-6}M_p$,
where $\phi \sim \mathcal{O}(10) M_p$.


Having established the isotropic
inflationary trajectory, we can now ask how this is modified in the
presence of a gauge field. Using the Einstein equations outlined
above, we see that the equation for accelerated expansion (in its
entirety) becomes
\begin{equation}\label{eq:accelerated_expansion}
\frac{\ddot{a}}{a} = -2 \dot{\sigma}^2 - \frac{\dot{\phi}^2}{3
M_p^2} + \frac{1}{3 M_p^2}\left(V + \frac{1}{b}\left\lbrack
1-\sqrt{1+b P_A^2 e^{-4\alpha-4\sigma}} \right\rbrack \right),
\end{equation}
where inflation occurs when we neglect the kinetic terms compared to
the potential. One can further see that inflation soon ends
\emph{if} the gauge field contribution dominates the scalar
potential term in the slow roll regime - suggesting that the
inclusion of a gauge field term will
always spoil inflation.
However in order to retrieve a more
detailed appraisal, it is important to note that the gauge field is
actually decoupled from the scalar sector, i.e., there is no source
term in the equation of motion for $\phi$ - so the gauge field
backreaction can only arise in the Klein-Gordon expression through
the definition of the Hubble parameter.
Let us then establish how a gauge
field can modify inflation. For
practical (computational) purposes, we will assume the slow roll
approximation holds, and that
\emph{initially}
 $\sigma, \dot{\sigma} \sim 0$ to leading
order. Consequently, due to the
special algebraic properties of the scalar sector, we can
again employ the Hamilton-Jacobi
formalism to solve for the field trajectory
\begin{equation}
\phi'(\alpha) = -\frac{M_p^2 \partial_{\phi }V(\phi)}{V(\phi)-
b^{-1} + b^{-1} \sqrt{1+bP_A^2 e^{-4\alpha}}},
\end{equation}
valid for any scalar potential. Note that
herein a prime denotes derivative
with respect to $\alpha$. Let us consider the case with $p=2$, and
consider a perturbative expansion \footnote{This is actually all we
can do analytically, the full solution does however admit a numeric
solution} in $P_A$ . The resulting terms at leading order are
therefore
\begin{equation}
\phi(\alpha) \sim \sqrt{\phi_0^2 -4\alpha M_p^2} \left( \pm 1 +
\frac{P_A^2 e^{-\phi_0^2/M_p^2}}{2m^2(\phi_0^2 - 4\alpha M_p^2)}
\left(\pm \Gamma\left \lbrack-\frac{1}{4},
-\frac{\phi_0^2}{M_p^2}\right\rbrack \mp
\Gamma\left\lbrack-\frac{1}{4}, \frac{4 \alpha
M_p^2-\phi_0^2}{M_p^2}\right\rbrack + \ldots \right)\right),
\end{equation}
where we have fixed the constant of integration to ensure that
$\phi(0) \sim \phi_0$ and $\Gamma$ are gamma functions \cite{Grad}.
The choice of sign arises from the square-root solution at leading
order, although we will assume (for simplicity) that $\phi$ is
non-negative over the inflationary domain. Numerically we do not
have to assume a perturbative expansion in the gauge field. However
we see that the solution is remarkably insensitive to $b$. Indeed
for fixed mass and field density, the scalar trajectories are almost
identical even if $b$ varies over factors $\mathcal{O}(10^4)$.

Regarding the evolution of spatial
anisotropy, under the herein assumptions, let us consider the
second of the Einstein equations (\ref{2.8}), in the limit where we
can drop the $\ddot{\sigma}$ term. In the perturbative
setting we have just discussed, we
can re-write this equation for small $\dot{\sigma}^2$ as
\begin{equation}\label{eq:anisotropies}
\frac{\dot{\sigma}}{\dot{\alpha}} \sim \frac{P_A^2}{3 V(\phi)}
\end{equation}
for an arbitrary potential. Note that this automatically implies
that if $V$ decreases during inflation, anisotropies are
automatically increasing. Specialising to our solution for the
quadratic case, we see that the back-reacted scalar field has higher
order correction terms in $P_A$ and therefore we see that the
classical potential  drives the
anisotropies to increase.

Because the gauge and scalar sectors are decoupled, the scale factor
determines the subsequent evolution
of electromagnetic energy. Indeed we can write the general energy
density as
\begin{equation}
\rho_A \sim \frac{1}{b}\left(\sqrt{1 + b P_A^2 e^{-4\alpha-4\sigma
}}-1 \right),
\end{equation}
which suggests that at late times, the gauge field contribution to
the energy density vanishes - although could still be important if
it dominates the scalar contribution. In order to
control this, one should arrange
for the gauge field contribution to be sub-leading during inflation.
We then consider the ratio of the two competing energy densities in
the slow-rolling phase
\begin{equation}\label{eq:Rdef}
\mathcal{R} = \frac{\rho_A}{\rho_{\phi}} \simeq \frac{\sqrt{1+bP_A^2 e^{-4\alpha-4\sigma}}-1}{bV(\phi)},
\end{equation}
which should be constant during inflation.
Perturbative expansion of the ratio results in the general expression
\begin{equation}
\mathcal{R} \simeq \frac{P_A^2 e^{-4\alpha - 4\sigma}}{2V(\phi)} + \ldots
\end{equation}
which can only be constant when $V
(\phi) \sim C e^{-4\alpha-4\sigma}P_A^2$ due to the cancellation
between powers of $b$. The constant $C$ is dimensionless in the
expression above. Clearly this solution does not correspond to the
power law potential due to the relation between $\alpha$ and $\phi$
through the Einstein equation. This means that we cannot
simultaneously find inflating trajectories with a constant value of
$\mathcal{R}$ for the power law solution.

If we demand that the perturbative bound on $\mathcal{R}$ is the
most important consideration, we must have $V \sim C e^{-4\alpha} P_A^2$
in the isotropic limit. The task now is to reconstruct the scalar
potential as a function of $\phi$. Note that this means the energy
densities of scalar and gauge sectors are now proportional with $C$
determining which is larger. Solutions with $C>1$ imply that the
scalar energy density is still dominant, whilst the gauge field
dominates for $C < 1$. Referring to
(\ref{eq:accelerated_expansion}), one sees that accelerated
expansion of the universe requires $C > 1$ - therefore the scalar
dominance should be considered the physical solution. Solving the
Hubble expression for the scale factor yields;
\begin{equation}
e^{2\alpha} \sim  1+ \frac{2 P_A t}{M_p}
\left(\frac{1+C}{3}\right)^{1/2},
\end{equation}
and then one can solve the scalar field equation to obtain
\begin{equation}
\phi \sim \phi_0 + M_p\sqrt{\frac{C}{1+C}}\ln \left( \frac{1}{1+A_1
t}\right) \hspace{0.5cm},  \hspace{0.5cm}  A_1 \equiv  \frac{2
P_A}{M_p} \sqrt{\frac{1+C}{3}},
\end{equation}
which suggests the scalar potential must be of the form
\begin{equation}
V \sim C P_A^2 \exp \left( \frac{4
(\phi-\phi_0)}{M_P}\sqrt{\frac{1+C}{C}}\right),
\end{equation}
which is a decreasing function of $\phi$ - because $\phi$ is initially chosen to be large.
Note that this simplifies
further if we assume the limit $C
>> 1$. However this exponentially decaying potential is proportional
to the gauge field term (which is already small), and moreover the
functional form does not admit a slow-roll inflationary solution
because the potential is far too steep. In fact, calculation of the
slow-roll parameters give
\begin{equation}
\epsilon \sim \frac{8(1+C)}{C}, \hspace{1cm} \eta = 2 \epsilon,
\end{equation}
which clearly do not allow for these inflating trajectories. This
is not a surprising result. We have tried to constrain the theory to
ensure that the total energy density remains constant during
inflation - forcing the scalar and gauge field energy densities to
be proportional to one another. As long as the scalar sector
dominates, expansion occurs. However this constraint on the scalar
potential reduces the Hubble friction in the equation of motion,
thus the scalar field cannot drive a
slow roll inflationary phase whilst satisfying the observational
bounds. Thus although the universe expands, scalar field driven
(slow roll) inflation is
inconsistent. The simplest way to resolve this issue is to increase
the number of scalar fields, ensuring that they all follow the same trajectory. This
is known as Assisted Inflation
\cite{Liddle:1998jc, Kanti:1999vt, Kanti:1999ie, Copeland:1999cs}, whereby the combined effect of $N$ scalars
will be to increase the Hubble friction term in the equation of motion, ensuring
that the overall centre of mass mode will follow an inflating trajectory.

Imposing the constraint (\ref{eq:Rdef}) on the energy densities is
too strong as it stands. A way to modify the theory is to introduce
some kind of scalar source term, ensuring that this constraint is
automatically satisfied. We will address this in the next section.
\subsection{$b(\phi)$ Coupled solution}
\label{b-phi}

\indent

Let us now consider a modification of the theory where we promote
the BI parameter to be a function of the inflaton $b(\phi)$. This
induces a non-trivial coupling to the gauge field, thereby allowing
the gauge field to act as a source
term as follows:
\begin{equation}
\ddot{\phi} + 3\dot{\alpha}\dot{\phi} + \partial_{\phi} V +
\frac{\partial_{\phi} b}{b^2}\left(1-\sqrt{1+b P_A^2
e^{-4\alpha-4\sigma}} \right).
\end{equation}
For inflating trajectories the gauge field term
eventually vanishes and we can
solve the field equation once we specify the scalar potential as
before. We will consider the quadratic potential as an explicit
example. Because of the coupling to $\partial_{\phi} \ln(b)$ in the
field equation, the backreaction on the scalar sector can occur
\emph{before} the gauge field energy density becomes comparable to
the scalar energy density. Therefore we can solve the back-reaction
problem at the level of the scalar field equation, and neglect the
gauge field energy density in the Hubble parameter. Perturbatively
we can then find the back-reacted Hamilton-Jacobi equation
\begin{equation}\label{eq:scalar_soln_2}
\phi' (\alpha)  \sim -\frac{2 M_p^2}{\phi}\left(1-\frac{P_A^2
e^{-4\alpha}}{m^2 \phi^2}\left(1+\frac{\partial_{\phi}
b}{b}\frac{\phi}{2} \right) \right),
\end{equation}
which is clearly sensitive to a term $\partial_{\phi} \ln b$ on the
right hand side. Let us consider a
few specific cases:

\begin{itemize}
\item If $\partial_{\phi} \ln b$ vanishes, or if $b$ takes the form $b
\sim A \phi^n$,  then the source term simplifies dramatically.
\begin{itemize}
\item In particular, there is  a critical value of $b$ for which the
back-reaction term vanishes, and that is when $b \sim b_0
\phi_0^2/\phi^2$. This last condition is particularly interesting
because it implies that at late times (when $\phi \to 0$) we recover
the Einstein-Maxwell theory.
\item Inflationary trajectories with $b
\sim \phi^{-2}$ are therefore unaffected by the additional
electromagnetic energy density (for small $P_A$ and with quadratic
potential), however it may lead to an interesting contribution
during reheating.
\end{itemize}
\item Consideration of the energy density ratio ${\mathcal{R}}= \frac{\rho_A}{\rho_\phi}$ (cf. (\ref{eq:Rdef})),
i.e., meaning to find a constant ratio and where the energy density
of the gauge field is sub-dominant, yields the following constraint
upon the Born-Infeld parameter
\begin{equation}\label{eq:b_soln_1}
b(\phi) \sim \frac{b_0}{(\mp 1 + \sqrt{1+b_0 P_A^2})^2}\left(b_0
P_A^2 e^{(\phi^2-\phi_0^2)/M_p^2} - 2\left (\sqrt{1+b_0 P_A^2} \mp 1
\right) \right),
\end{equation}
where $b_0$ is a constant set by the initial conditions on the
inflaton field. Again we can only obtain a perturbative
solution for the scalar field
\begin{equation}\label{eq:b_soln}
\phi \sim \sqrt{\phi_0^2-4\alpha M_p^2}\left(1 +
\frac{e^{-\phi_0^2/M_p^2} P_A^2}{2 m^2(\phi_0^2-4\alpha
M_p^2)}\left(\Gamma\left \lbrack 0, -\frac{\phi_0^2}{M_p^2} \right
\rbrack+ \Gamma \left \lbrack 0, \frac{4 \alpha M_p^2 -
\phi_0^2}{M_p^2} \right \rbrack \right) + \ldots \right),
\end{equation}
which is similar to the case of constant $b$ discussed earlier. This
again suggests that the theory is insensitive to the Born-Infeld
parameter. Numerically we can integrate the full scalar field
equation as before once we specify the scalar dependence of
$b(\phi)$. The resulting plots  confirm that the solution is indeed
insensitive to $b(\phi)$.
\end{itemize}

Concerning the anisotropies
generated during this regime, the expression is the same as that
derived in (\ref{eq:anisotropies}) and therefore we need not write
it again. The scalar field solution contributes terms of higher
order in $P_A$ and thus we find the same result as in the decoupled
case.


The back-reaction on the scalar field is important cosmologically, however, because the
number of e-foldings is determined through the expression
\begin{equation}
\mathcal{N} = \int \frac{d \phi}{\phi(\alpha)'}.
\end{equation}
Let us assume that $b \sim A \phi^n$ for now, which is the simplest
non-trivial case we can consider (aside from $n = -2$). At leading
order in a perturbative expansion we obtain the following solution
for the scalar field at horizon crossing from (\ref{eq:scalar_soln_2})
\begin{equation}
\phi_{*}^2 \sim 2M_p^2(1+2\mathcal{N}) + \frac{P_A^2}{m^2}
e^{-\phi_0/M_p^2}\left(1+\frac{n}{2}\right)
\left(Ei(2)-\gamma-2M_p^2(1+2\mathcal{N}) \right) + \ldots ,
\end{equation}
where $\phi_0$ is the initial value of the scalar field, and
inflation ends at $\phi = \sqrt{2}M_p$
and $Ei$ is the expontential integral function \cite{Grad}. We have also neglected
higher order terms, and work in a regime where $\phi^2/M_p^2 >>
\ln(\phi^2/M_p^2)$ which allows us to neglect the logarithmic
contributions. This latter approximation is physically well
justified over the field domain.

With this expansion we can estimate the cosmological perturbations
at horizon crossing which we then write in terms of physical
observables. It is convenient to define the following function
$\xi(\mathcal{N}) \equiv  2 + \gamma + 4\mathcal{N}-Ei(2)$,  which
is positive definite for all $\mathcal{N}
> 1$. Then we find at leading order
\begin{eqnarray}
n_s &\simeq& 1-\frac{4}{1+2\mathcal{N}} - \frac{(2+n)
\xi(\mathcal{N})}{(1+2\mathcal{N})^2} \left( \frac{P_A^2
e^{-\phi_0^2/M_p^2}}{m^2 M_P^2} \right), \\
r&\simeq& \frac{16}{1 + 2\mathcal{N}} \left\lbrace 1+ \frac{4
(2+n)\xi(\mathcal{N})}{(1+2\mathcal{N})} \left( \frac{P_A^2
e^{-\phi_0^2.M_p^2}}{m^2 M_p^2}\right) \right\rbrace, \\
\mathcal{P}_s^2 &\simeq& \frac{(1+2\mathcal{N})^2 m^2}{24 \pi^2
M_p^2} \left\lbrace 1 - \frac{(2+n \xi(\mathcal{N}))}{2
(1+2\mathcal{N})}\left( \frac{P_A^2 e^{-\phi_0^2}/M_p^2}{m^2
M_p^2}\right) \right\rbrace, \\
\mathcal{P}_t^2 &\simeq&
\frac{2m^2(1+2\mathcal{N})}{3 \pi^2 M_p^2}\left \lbrace 1 -
\frac{(2+n) \xi(\mathcal{N})}{4 (1+2\mathcal{N})} \left(\frac{P_A^2
e^{-\phi_0^2/M_p^2}}{m^2 M_p^2} \right)\right\rbrace.
\end{eqnarray}
Using the WMAP $7$ year bound \cite{Larson:2010gs} on $n_s$, which
satisfies $n_s \sim 0.963 \pm 0.012$ \cite{Bennett:2010jb,
Larson:2010gs} and using $\mathcal{N}=60$ to denote horizon
crossing, we find that the following constraint on the backreaction
parameter $Q$
\begin{equation}
Q = \frac{P_A^2 e^{-\phi_0^2/M_p^2}}{m^2 M_P^2} \sim
\frac{0.24^{+0.74}_{- 0.24}}{(2+n)},
\end{equation}
where $Q=0$ corresponds to the case of zero gauge field. Now we note
that the tensor-scalar ratio appears to be an increasing function of
$n$, therefore the back-reaction term will lead to larger than
observed values of $r$ unless $n$ is negative. This means that for
successful inflation we can place a rough bound
on the physically allowed values of
$n$. We illustrate this in  Figure
1, shown  below over a small range of $n$ around $n\sim -2$ for the
above bounds on the expansion parameter.
\begin{figure}[htp]
\begin{center}
\includegraphics[width=0.7\textwidth]{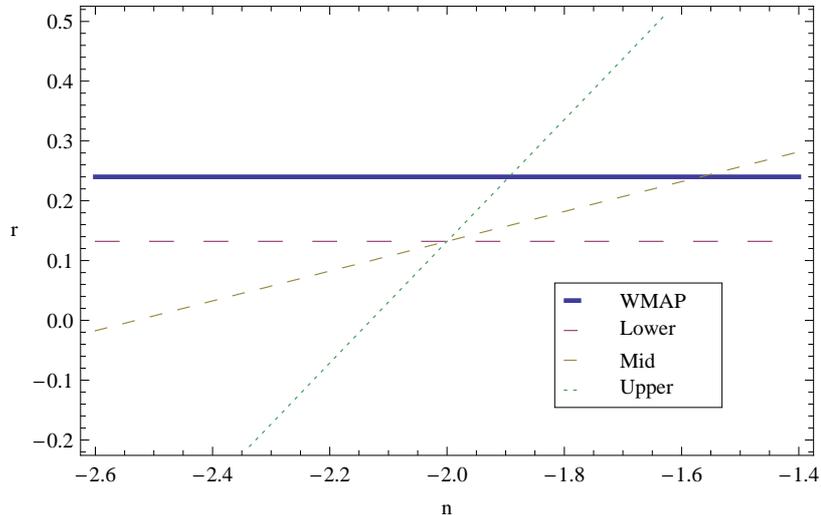}
\caption{Plot of the scalar-tensor ratio as a function of power law
index $n$. The WMAP bound implies that only values of n $\in (-2.4,
-1.6)$ are physically acceptable. The lower bound arises because a physical $r$ must be non-negative.}
\end{center}
\end{figure}
As one can see, the lower bound implies a vanishing of the $P_A^2$
term and therefore a constant value of $r$ which is below the WMAP
bound. This is to be expected, since this is just canonical
inflation. For non-zero values of the $P_A$ term up to the maximal,
we see that there is a small window where $r$ satisfies the WMAP
bound, and is positive definite.

Values of $n$ outside the identified range are incompatible with observation, or non-physical therefore
should be ruled out as viable candidates for inflation. As expected from our earlier
observation of the Hamilton-Jacobi equation, the theory with $n=-2$ is indistinguishable
from inflation without gauge field (at leading order) and therefore easily satisfies
the WMAP constraint.

One could examine the entire parameter space to classify the
inflating trajectories in terms of the BI functions, however this is
not the main focus of this paper, and we leave the full parameter
space evaluation for future work. We comment briefly on the case of
exponential coupling $b(\phi) \sim A e^{a \phi^2}$,
$a$ is a constant,  (cf.
\ref{eq:b_soln_1}) which may be of interest. In this case the scalar
field admits the following perturbative solution
\begin{equation}
\phi(\alpha) \sim \sqrt{\phi_0^2 - 4 \alpha M_p^2}\left(1-\frac{e^{-\phi_0^2/M_p^2} P_A^2}{2 m^2 (\phi_0^2 - 4\alpha M_p^2)}\left(Ei\left(\frac{\phi_0^2}{M_p^2}\right)
+ Ei\left(\frac{\phi_0^2-4\alpha M_p^2}{M_p^2}\right) -2 \alpha M_p^2 e^{-4\alpha + \phi_0^2/M_p^2} \right) \right)
\end{equation}
where $Ei(x)$ is the ExpIntegralEi(x) function \cite{Grad},  and the field satisfies the usual boundary
condition that $\phi(0) = \phi_0$. Note that this solution differs
from that in (\ref{eq:b_soln}) through the use of the Ei function.
We will now turn to a related issue, examining how gauge fields
interact with the DBI theory of $D3$-branes.

\section{D-Brane theory}
\label{DBI}

\indent

As briefly mentioned in sections 1
and 2, the non-linearity type of structure present in the
Born-Infeld frameworks turns out to provide the correct low energy
description of $Dp$-branes in string theory. In this
context  a class of models within
type IIB string cosmology, more often referred to as DBI-cosmology
has been considered
\footnote{Inflation in these models is driven by the motion of a
probe $D3$-brane through a warped geometry. The warped geometry is
typically taken to be a solution of the ten-dimensional field
equations, which is then `glued' to a compact manifold - and the
probe brane is localised within the geometry far away from this
gluing region. Since inflating trajectories are dependent on the
particular warped background, we will consider instead  a more
phenomenological approach in this section - making reference to
relevant string theory solutions as they arise}.
Notwithstanding their interest, the
simplest models of DBI-cosmology are plagued by several issues which
disfavor them as inflationary candidates \cite{Leblond:2008gg,
Kobayashi:2007hm, Huston:2008ku}. More complex models can be
constructed which circumvent these issues, but they all appear
sensitive to supergravity back-reaction in the relativistic limit
\cite{Kobayashi:2007hm, Becker:2007ui, Berndsen:2009ww}. All these
particular models make use of internal spaces that are
asymptotically AdS, and it has not
yet been established whether more general backgrounds are more
viable. Recently  the non-trivial
coupling between dilaton and gauge field has been exploited in
\cite{Avgoustidis:2008zu} through the use of Wilson lines: Inflation
in such a model appears to avoid several of the pathologies of
previous DBI inflation, and
suggests that such non-trivial couplings are important to realise an
inflationary phase. Since such non-trivial couplings are generated
through non-AdS backgrounds, our aim
in what follows is to consider such
geometries in the hope that they can be embedded into the full
string theory. Throughout this
section we will discuss how some simpler (Lagrangian) configurations
can be of interest. We will show how we can compute modifications on
the scalar field dynamics (and hence, the inflationary universe),
given constraints on the electromagnetic energy density. More
concretely,  we will investigate specific limits of the  equations
of motion; Namely the non-relativistic and relativistic limit (cf.
subsections 3.1 and 3.2, respectively). These will allow us to
establish backreaction situations of gauge fields on the inflaton
and possibly identify cosmological implications, namely
observational. The feature to stress is that the allowed couplings
induce herein a much richer class of dynamical possibilities than
the content in section 2 (A brief discussion on generating magnetic
seed fields suitable for these limits is first presented in Appendix
A.).

Let us start with the following effective theory for DBI cosmology
for a single $D3$-brane (which is in Einstein frame), where we
include a scalar potential which does not come from the world-volume
calculation but could arise from a supergravity F-term,
\begin{equation}
S = \int d^4 x \left( -F_1(\phi)\sqrt{-{ \rm det}(G_{\mu \nu}+
F_3(\phi)F_{\mu \nu})} + q \sqrt{-g} F_1(\phi) -
\sqrt{-g}V(\phi)\right),
\end{equation}
where the induced metric is
\begin{equation}
G_{\mu \nu} = g_{\mu \nu} + F_2(\phi)\partial_{\mu} \phi \partial_{\nu} \phi
\end{equation}
and $q$ is a measure of the RR-charge carried by the brane, where
$q=1$ for a BPS $D3$-brane, $q=-1$ for a $\bar{D}3$-brane and $q=0$
corresponds to non-BPS branes such as employed in models of tachyon
dynamics. We will assume that $F_1$ is non-negative for simplicity.
Negative tension objects must exist for a string compactification to
be consistent, but we will not consider them in this particular
paper. The bulk metric $g_{\mu \nu}$ is assumed to be a solution of
type IIB string theory in Einstein frame. Our ansatz (\ref{B-I}) is,
in fact, general enough to extend to these backgrounds which
typically have a non-trivial dilaton profile - which is important
because the determinant contains additional powers of the dilaton,
coupling to the Maxwell tensor in Einstein frame. Backgrounds
falling into this class can be found in \cite{Bigazzi:2009bk,
Nunez:2010sf}.


One can calculate the determinant at leading order and therefore can write the above action as follows
\begin{equation}
S = \int d^4 x \left(
-F_1(\phi)\sqrt{-G}\left(1+\frac{1}{4}F_3^2(\phi) F_{\mu \nu} F^{\mu
\nu}\right) + \sqrt{-g} (qF_1(\phi)-V(\phi)) \right),
\end{equation}
where indices are raised and lowered with the full metric $G_{\mu
\nu}$ rather than $g_{\mu \nu}$. Note that we are neglecting any
couplings to the axion, which in this theory could come in the form
$c_0 F \wedge F$. The $F_i$ are functions of the scalar embedding,
arising from placing the $D3$-brane in the non-trivial geometry. The
factors of the brane tension are contained in the $F_1$
parameter. This action appears
superficially similar to that of Einstein-Maxwell theory, which
typically has a coupling between scalar and gauge sectors
as $b^2(\phi) F_{\mu \nu} F^{\mu
\nu}$. In the D-brane theory we see that this coupling is fixed to
be $F_1 F_3^2$, which potentially allows for much richer solutions
than the function $b(\phi)$. Note however that the Maxwell tensors
also couple to the inflaton through the term $\sqrt{-G}$, which
contain powers of $\dot{\phi}^2$. For dynamic solutions, this
coupling explicitly breaks conformal invariance. Therefore even for
theories with $F_1 F_3^2 = 1$, the conformal structure of the gauge
field sector is still broken \cite{Bamba:2008my}.


The modified Klein-Gordon equation is much more complicated, but
takes the following form
\begin{eqnarray}
0&=& \left(1+\frac{F_3^2}{4}F_{\mu \nu} F^{\mu \nu} \right)\left \lbrace F_1F_2' \dot{\phi}^2(1- 4 e^{3 \alpha})-2 F_1'\left \lbrack 1-F_2 \dot{\phi}^2 + \frac{e^{3\alpha}F_1 F_2 \dot{\phi}^2}{(1-F_2 \dot{\phi}^2)}\right \rbrack -6e^{3\alpha}\dot{\alpha}F_1F_2 \dot{\phi}\right \rbrace \nonumber \\
&-& 2 F_1 F_2 \ddot{\phi} e^{3\alpha} \left(1+\frac{F_3^2}{4}F_{\mu \nu} F^{\mu \nu} \right) - F_1F_3 F_3'\frac{F_{\mu \nu}F^{\mu \nu}}{2\sqrt{1-F_2 \dot{\phi}^2}}\left( 1- F_2 \dot{\phi}^2 + e^{3\alpha} F_3 \dot{\phi}^2\right) \nonumber \\
&+& qF_1' - V' - \frac{e^{3\alpha}F_1 F_2 F_3^2 \dot{\phi}}{4
\sqrt{1-F_2 \dot{\phi}^2}} \frac{\partial}{\partial t}(F_{\mu \nu}
F^{\mu \nu}),
\end{eqnarray}
where a dot denotes a time derivative, and primes denote
herein scalar field derivatives.
For backgrounds which are asymptotically $AdS$, the string and
Einstein frames coincide - leading to a theory with
a trivial dilaton (at leading
order). Canonical examples of such theories include;
\begin{itemize}
\item $F_1 = F = \phi^4/ \lambda, F_2 = f, F_3 = 1/\sqrt{F}, q = \pm 1$ for the AdS
solution;
\item $F_1 = V(\phi), F_2 =1, F_3 = \lambda, q=0$ for the tachyon
solution \cite{Sen:2002nu, Sen:1999md, Kluson:2000iy,
Garousi:2000tr}
\end{itemize}
We must now consider the Einstein equations for such a
configuration. These will generally be complicated by the presence
of the induced metric $G$ rather than the space-time metric $g$.
Using the variational identity
\begin{equation}
\delta G^{ab} = \delta g^{ac} (\delta ^c_b +
g^{ce}h_{eb})^{-1}-\delta g^{\mu \nu} g^{ac}h_{\nu \sigma} (\delta
^{\mu}_b + g^{\mu x}h_{xb})^{-1}(\delta^c_{\sigma} +
g^{c\rho}h_{\rho \sigma})^{-1},
\end{equation}
one can calculate the corresponding energy momentum tensor to be
\begin{eqnarray}
T_{\mu \nu} &=& -\frac{1}{4}\sqrt{1-F_2 \dot{\phi}^2}F_1 F_3^2 F_{ab}F^{ab} \left(G_{\mu b}(\delta ^{\nu}_b + g^{\nu e}h_{eb})^{-1} - G_{ab}g^{ac}h_{\nu \sigma}(\delta^{\mu}_b + g^{\mu e}h_{eb})^{-1}(\delta^c_{\sigma} + g^{ce}h_{e\sigma})^{-1}\right) \nonumber \\
&+& \frac{1}{2} F_1 F_3^2 \sqrt{1-F_2 \dot{\phi}^2} \left(F_{\mu b}F_{cd}G^{bd}(\delta^{\nu}_c + g^{\nu e}h_{ec})^{-1}+ F_{a \mu}F_{cd}G^{ac}(\delta^{\nu}_b
+ g^{\nu e}h_{eb})^{-1} \right) \nonumber \\
&-& \frac{1}{2} \sqrt{1-F_2 \dot{\phi}^2}h_{\nu \sigma}(\delta^l_{\sigma} + g^{le}h_{e\sigma})^{-1}\left(G^{bd}g^{al}(\delta^{\mu}_c + g^{\mu f}h_{fc})^{-1} + G^{ac}g^{bl} (\delta^{\mu}_d + g^{\mu k}h_{kd})^{-1} \right) \nonumber \\
&+& F_1 \sqrt{1-F_2 \dot{\phi}^2}\left(-G_{\mu b}(\delta ^{\nu}_b + g^{\nu e}h_{eb})^{-1}+G_{ab}g^{ac}h_{\nu \sigma}(\delta^{\mu}_b + g^{\mu e}h_{eb})^{-1}(\delta^{c}_{\sigma} + g^{c f}h_{f\sigma})^{-1} \right) \nonumber \\
&+& g_{\mu \nu}(q F_1 - V),
\end{eqnarray}
where we have defined $h_{ab} = F_2 \partial_a \phi \partial_b \phi$
for simplicity. Since the DBI action has a well-defined relativistic
limit, we choose to define $\gamma \equiv (1-F_2
\dot{\phi}^2)^{-1/2}$ as the analog of the `gamma' factor in special
relativity. The resulting Einstein equations arising from the
Bianchi metric, can then be written as follows
\begin{eqnarray}
\dot{\alpha}^2 - \dot{\sigma}^2 &=& \frac{1}{3 M_p^2}\left(
F_1(\gamma-q) +V + \frac{F_1 F_3^2 \gamma^3
\dot{A_x^2}e^{-2\alpha+4\sigma}}{2}\right), \label{3.7} \\
\ddot{\sigma}+3\dot{\alpha} \dot{\sigma} &=&
\frac{e^{-2\alpha+4\sigma}F_1 F_3^2 \dot{A_x^2}\gamma(\gamma^2+1)}{6
M_p^2},  \label{3.8}\\
2\ddot{\alpha}+3\dot{\alpha}^2+3\dot{\sigma}^2 &=&
\frac{1}{M_p^2}\left( \frac{F_1(1-\gamma q)}{\gamma} + V -
\frac{e^{-2\alpha+4\sigma }F_1 F_3^2 \dot{A_x^2}\gamma
(2\gamma^2-1)}{6}\right) \label{3.9}
\end{eqnarray}
and one finds that the equation for an accelerating universe is given by
\begin{equation}
\frac{\ddot{a}}{a} = -2 \dot{\sigma}^2 + \frac{1}{3
M_p^2}\left(\frac{F_1(3-2\gamma q - \gamma^2)}{2 \gamma} + V -
\frac{e^{-2\alpha+4\sigma} F_1 F_3^2
\dot{A_x^2}\gamma(3\gamma^2-1)}{4} \right).
\end{equation}
From the right hand side of this equation one sees that the energy
density of the vector field is important when considering
inflationary dynamics. Indeed we can identify a critical value of
the gauge field which allows for acceleration (isotropic limit):
\begin{equation}\label{eq:constraint}
\dot{A_x^2} < \frac{4 e^{2\alpha-4\sigma}}{F_1 F_3^2 \gamma (3
\gamma^2-1)}\left(V - \frac{F_1(\gamma^2 + 2\gamma q - 3)}{2 \gamma}
\right).
\end{equation}
Note that in DBI inflation, the scalar potential dominates the energy density even
for relativistic rolling. It can then be seen that the term on the right hand side is
a decreasing function as one approaches the relativistic limit. This reduces the
solution space for $\dot{A}_x^2$ so that it approaches zero as $\gamma \to \infty$.

Finally upon variation of the
action we find the coupled Maxwell equation
\begin{equation}
\partial_{\mu} \left( \sqrt{-G} F_1(\phi) F_3^2(\phi) F^{\mu
\nu}\right)=0, \label{3.12}
\end{equation}
which mixes the inflaton with the gauge field. Inserting our ansatz
from section 2.1 we find that the
equation becomes
\begin{equation}
\ddot{A}_x + \dot{A}_x
\left(\dot{\alpha}+4\dot{\sigma}+\frac{F_1'}{F_1}\dot{\phi}+\frac{2
F_3'}{F_3}\dot{\phi} + \frac{F_2'\dot{\phi}^3}{(1-F_2
\dot{\phi}^2)}+ \frac{2F_2 \dot{\phi}\ddot{\phi}}{(1-F_2
\dot{\phi})} \right)=0. \label{3.13}
\end{equation}
Solutions to the Maxwell equation include the following - similar to
that in the BI-theory
\begin{equation}
\dot{A}_x \simeq -\frac{P_A e^{-\alpha-4\sigma}}{ \gamma F_1 F_3^2},
\label{3.14}
\end{equation}
where we denote the constant of integration as $P_A$.
In this case $P_A$ is a measure of
the charge carrier density on the world-volume because the gauge
field arises from excited states of open $F$-strings that end on the
brane.

In the non-relativistic limit $(\gamma \sim 1)$, the scalar
potential will effectively dominate the right hand side of the
constraint equation (3.11). Therefore for a sufficiently `large'
potential, this condition can easily be satisfied. In the
ultra-relativistic limit, the right hand side is proportional to $
\gamma^{-3}(V-\gamma F_1/2)$ which may not be very large without
fine-tuning, thereby making the constraint equation (3.11) very
difficult to satisfy. Clearly the gauge field will have the most
dramatic effect on relativistic inflation, but may also play an
important role in the non-relativistic limit. We can again, like in
subsection 2,  adapt the constraint on the energy density during
inflation, in the form of a  ratio of energy densities to  be
approximately constant (cf., e.g., (2.17)). In the case of the DBI
theory, the new expression for the ratio takes the form
\begin{equation}\label{eq:R_DBI}
\mathcal{R} = \frac{P_A^2 \gamma e^{-4\alpha-4\sigma}}{2 F_1 F_3^2
(F_1(\gamma-q) + V )},
\end{equation}
which clearly relates the various parameters in the model. Examining
the ratio (\ref{eq:R_DBI}) for sufficiently small values of
$\dot{\phi}$ - and considering the critical limit where the gauge
field energy density is almost constant during inflation - we find
the constraint
\begin{equation}\label{eq:condition_1}
F_1 F_3 ^2 \sim e^{-4  \alpha},
\end{equation}
which is the generalised extension of the known results from
Einstein-Maxwell theory. To further assist in understanding the role
of the gauge field, we use the
parameter $c$ (cf. Appendix A),
which conveys the deviation from the above result. More precisely,
the solution we must consider is $F_1 F_3^2 \sim e^{-4 a c}$ - which
is the most general expression. For $c > 1$, the gauge field terms
will rapidly come to dominate the dynamics at late times.


Before proceeding and specializing with different settings of our
DBI framework, let us indicate that the anisotropies are governed
(at leading order) by equations (\ref{3.7}), (\ref{3.8})  above.
Under the assumption that they are small, and obey slow-roll
behaviour, we can immediately write down the relevant
Hamilton-Jacobi equation which describes their dynamics
\begin{equation}\label{eq:anisotropies-a}
\frac{d \sigma}{d \alpha} \sim \left( \frac{P_A^2}{F_1 F_3^2}\right)
\left(\frac{1+\gamma^2 }{6\gamma}\right)
\frac{e^{-4\alpha}}{F_1(\gamma-q)+V}.
\end{equation}
In the non-relativistic limit, the $\gamma$ terms reduce to a
constant (given by $1/3$), and the magnitude is then set by
whichever term dominates the scalar energy density. For ultra
relativistic motion, the middle term appears to be linear in
$\gamma$, and if the scalar potential is dominant this suggests that
the anisotropies may be large. However if the $F_1 \gamma$ term
dominates the energy density, then the $\gamma$ factors actually
cancel - and the anisotropies are then determined by the new term
$F_1^2 F_3^2$.

\subsection{Non-Relativistic limit}
\label{non-rel}

\indent

The non-relativistic limit of the theory emerges when $\gamma \sim 1
+ \frac{1}{2} F_2 \dot{\phi}^2 +\ldots$. Although this is a modified
version of slow-roll inflation, there is a non trivial coupling to
the gauge field and therefore one may expect back-reaction to be an
issue. Note that for $q=1$ the effective potential $V_{eff} =
V(\phi) + F_1(\phi)(1-q)$ reduces to the scalar potential due to
supersymmetry, and when $q=0$ we have a purely non-BPS system -
which also allows for $V=0$ and
the dynamics are dictated by $F_1$ alone.

Expanding the DBI Lagrangian to leading order yields a
corresponding effective field equation
\begin{equation}
\ddot{\phi} + 3\dot{\alpha} \dot{\phi} + \dot{\phi}\frac{\dot{Q}}{Q}
- \frac{W'}{Q}-\frac{\dot{\phi}^2}{2}\frac{Q'}{Q} = 0,
\end{equation}
where herein dots denote time
derivatives, primes are derivatives with respect to the scalar field
and we employ the following variable definitions
\begin{eqnarray}
Q &\equiv &  F_1 F_2 \left( 1 + \frac{F_3^2}{4}F_{\mu \nu}F^{\mu
\nu}
\right) = F_1 F_2 \left(1-\frac{x}{2}\right), \\
W&\equiv & F_1(q-1) -V - F_1 \frac{x}{2},
\end{eqnarray}
where $x \equiv - \frac{F_3^2}{2} F_{\mu\nu}F^{\mu\nu}$.  We have
not made any assumptions about the background geometry at this
stage, and this result is therefore quite robust. Let us study the
background evolution by starting
with  the isotropic inflationary limit and  initially set the gauge
field and $\ddot{\phi}$ terms to zero, in which case the above
equation of motion reduces to
\begin{equation}
3\dot{\alpha}\dot{\phi} + \frac{\dot{\phi}^2}{2}
\frac{\partial}{\partial \phi} \ln (F_1
F_2)-\frac{(q-1)}{F_2}\frac{\partial}{\partial \phi} \ln(F_1) +
\frac{V'}{F_1 F_2} \sim 0 ,
\end{equation}
which is reminiscent of a canonically coupled scalar field solution
- except that there is an additional driving term proportional to
$\dot{\phi}^2$. Note that BPS configurations ($q=1$ in our language)
simplify the equation of motion significantly since the second term
in (3.21) vanishes in this limit (with the usual assumptions about
regularity of the background function). The troublesome
$\dot{\phi}^2$ terms, which have no analog in canonical slow roll
models, will always contribute - unless we can consider limits where
$F_1 F_2$ is constant. This amounts to localising the solution on a
curve in parameter space, and is what we consider in this paper.
General solutions to the above equation will be explored in the
future.

\subsubsection{\bf $F_1F_2$ Constant}

With the assumption that $F_1 F_2$
is constant, the field equation becomes a modified version of the
canonical scalar field equation. We can immediately write down the
scale factor as a function of the inflaton using the Hamilton-Jacobi
formalism, for an arbitrary scalar potential
\begin{equation}
\alpha = -\frac{F_1 F_2}{M_p^2} \int \frac{V}{V'} d\phi.
\end{equation}
For a simple chaotic inflationary potential such as $\frac{1}{2} m^2
\phi^2$ theory, we can then see that the back-reaction constraint
(\ref{eq:condition_1}) implies the following general relationship
\begin{equation}
F_1 F_3^2 \sim \exp \left( \frac{F_1 F_2 c \phi^2}{M_p^2}\right).
\end{equation}
which dictates the functional form of $F_3$, given that $F_1$ is fixed.

We now proceed to include the gauge field terms in the modified Klein-Gordon equation
which we write as a
perturbative expansion in $x$, and takes the general form
\begin{eqnarray}
0 &\sim & 3\dot{\alpha}\dot{\phi} \left(1+\frac{2x}{3} \right) + \frac{\dot{\phi}^2}{2} \frac{\partial}{\partial \phi} \ln(F_1 F_2) + \frac{V'}{F_1 F_2}
\left(1+\frac{x}{2} \right) \nonumber \\
&+& \frac{(x-(x+1)(q-1))}{2 F_2} \frac{\partial}{\partial \phi}
\ln(F_1) + \frac{x(F_2 \dot{\phi}^2-2)}{2 F_2}
\frac{\partial}{\partial \phi} \ln(F_1 F_3),
\end{eqnarray}
although the $\dot{\phi}^2$ term in the last bracket in the
expression above is actually subleading in the non-relativistic
expansion - and can be neglected. Note that the
source term also appears to be
coupled to the scalar potential, which is due to the non-canonical
nature of the theory. To solve this equation, we use the energy
density constraint (\ref{eq:R_DBI}) to solve for $F_3$ under the
assumption of a quadratic scalar potential - since this is the
simplest analytic solution. One can solve this equation numerically,
however we present only analytic solutions in this paper - leaving a
more exhaustive analysis to future work:

\begin{description}

\item{(a)} Let us initially consider $F_1$ to be constant, which
(by assumption) implies that $F_2$
is also constant. Using the back-reaction condition to solve for
$F_3$ we can then find the solution to the scalar field equation in
the presence of the gauge field. At late times the solution can be
calculated to  converge to the following
\begin{equation}\label{eq:DBI_SOLN_1}
e^{-4\alpha-F_1 F_2 c \phi^2/M_p^2} \sim \frac{6 (c-1) F_1 m^2
M_p^2}{c P_A^2(6 F_1^2 F_2 c - m^2 M_p^2)},
\end{equation}
where the right hand side is clearly a constant which vanishes for $c=1$, therefore
the electromagnetic energy density will also be constant.
We see that $F_1 F_3^2 \sim 1/a^{2}$ in the regime where $P_A$ is constant,
but more importantly it highlights the fact that $c \sim 1$ as the back-reaction
becomes relevant.

For a canonically coupled field, there is attractor behavior since
the gauge field energy density exhibits tracking behaviour \cite{Kanno:2009ei}. In our case, due to the
non-canonical nature of the action, the source term in the field
equations also couples to the scalar potential. If we demand that
the scalar and gauge fields are effectively the same magnitude, so
that they are both source terms in the field equation, then we find
\begin{equation}
m^2 M_p^2 \sim \frac{xc (F_1 F_2)^2}{F_2 M_p^2}\left(1+\frac{x}{6} \right)
\end{equation}
at leading order.
Inserting this into the relation (\ref{eq:R_DBI}) we discover that
\begin{equation}
\mathcal{R} \sim \frac{M_p^2}{c \phi^2 (F_1 F_2)}.
\end{equation}
Given that inflation occurs for $\phi \sim \mathcal{O}(10) M_p$ in
this model, we find $\mathcal{R} \sim x 10^{-2}/(F_1 F_2)$, which
can be vanishingly small for sufficiently large $F_1 F_2$. This
ensures that the gauge field energy density is negligible when
compared to the scalar energy density and therefore we can
consistently neglect its contribution to the Hubble parameter. Note
that this is almost the same result as obtained in canonical models,
aside from the factor of $F_1 F_2$. This ensures that $\mathcal{R}
\sim x 10^{-2}/(F_1 F_2)$ is an attractor solution. If $\mathcal{R}$
is initially smaller than this, the behaviour of the scalar field
drives $\rho_A$ to increase (with $c>1$), reaching this value from
below. For configurations where this quantity is larger than the
attractor solution, the inflaton can climb back up the scalar potential
(due to the source term in the equation of motion). Thus $\rho_A$ is
forced to decrease rapidly, and the attractor is reached from above.
This confirms that even this non-canonical model we expect tracker
behaviour.

As far as the anisotropies are concerned, dropping the
$\ddot{\sigma}$ term in the anisotropic equation of motion
(\ref{3.8}), combined with the
back-reacted solution implies the following
\begin{equation}\label{eq:DBI_NR_1}
\frac{\dot{\sigma}}{\dot{\alpha}} \sim \frac{2 P_A^2}{3 m^2 \phi^2}
\frac{6 (c-1) F_1 m^2 M_p^2}{c P_A^2(6 F_1^2 F_2 c - m^2 M_p^2)},
\end{equation}
indicating that anisotropies increase during this accelerated phase
for $c>1$ and $6 c F_1^2 F_2 > m^2 M_p^2$. Of course, in order to
determine this result we assume a perturbative $\sigma$ as before.
Note that, as in the case of Born-Infeld theory, the increase in
anisotropy is determined by the scalar potential. Both solutions
(\ref{eq:anisotropies}) and (\ref{eq:DBI_NR_1}) increase like
$1/\phi^2$ - although in the latter case the anisotropies also
vanish when $c=1$.


\item{(b)} Consider now a regime of
solution space where the logarithmic terms dominate in the
back-reacted equation of motion - still assuming that $F_1 F_2$ is
constant. The solution for the scalar field in this instance becomes
\begin{equation}
e^{-4\alpha-F_1 F_2 c \phi^2/M_p^2} \sim \frac{e^{-4\alpha} m^2
M_p^2}{n F_1^2 F_2^2 c^2 P_A^2(1-e^{-4\alpha})+m^2 M_p^2 e^{F_1F_2 c
\phi_0^2/M_p^2}},
\end{equation}
where $n$ is an integer depending on whether $F_1 (n=1)$ or $F_3
(n=2) $ is fixed to be constant. Note that this
is similar to the result obtained in (\ref{eq:DBI_SOLN_1}) with the
additional dependence on the scale factor appearing on the right
hand side which ensures the solution decreases as a function of
time. The scale factor dependence arises precisely because of this
logarithmic running. The solutions are valid in a regime where the
scalar mass in the quadratic potential satisfies the following bound
\begin{equation}
m^2 << \frac{nxc F_1^2 F_2}{M_p^2}\left(1-\frac{x}{6} \right)
\end{equation}
where $n$ is defined as above, and when $n=2$ we must recall that
 $F_1$ depends on the inflaton. This bound can be satisfied
for a (small) region of solution space and is therefore a physical
solution. If the parameters are chosen so as to satisfy this bound,
then one can easily show that for
initially small  anisotropies, these  are actually decreasing
during the inflationary expansion - indicating that the isotropic
universe would be  a late time
attractor.



\item{(c)} It is also of interest to explore a solution branch
where we fix the functional form of $F_3$, but with the same
assumption that the logarithmic terms are dominant with respect to
the scalar potential. This should then modify the above result for
the scalar field. The ansatz we select is $F_3 \sim L\phi^a$  -
which allows us to find the analytic solution
\begin{equation}
e^{-4\alpha-c F_1 F_2 \phi^2 /M_p^2} \sim  \frac{m^2 M_p^2
e^{-4\alpha}}{F_1^2 F_2^2c^2 P_A^2(1-e^{-4\alpha})+m^2 M_p^2 e^{F_1
F_2 c\phi_0^2/M_P^2}},
\end{equation}
which is in fact identical to the solution above for $n=1$.
This suggests  that the anisotropies
are decreasing in this regime - since their overall magnitude is a
decreasing function of $\alpha$. Let us therefore consider a
solution where $F_1 \sim L \phi^a$ - which continues to fix $F_2$,
but now with $F_3$ as a constant. The resulting expression for the
back-reacted field equation is
\begin{equation}
\phi'(\alpha) \sim \frac{2 F_1 F_2 c P_A^2}{m^2} e^{-4\alpha-F_1 F_2 c \phi^2/2M_p^2}
\end{equation}
which admits a complicated solution of the form
\begin{equation}
e^{-cF_1 F_2 \phi^2/M_p^2} \sim \exp \left(2 {\rm Inverse
Erf}\left(i {\rm Erfi}\left \lbrack
\frac{\phi_0}{M_p}\sqrt{\frac{F_1 F_2 c}{2}}\right \rbrack +
\frac{i(F_1 F_2 c)^{3/2} P_A^2(1-e^{-4\alpha})}{m^2 M_p
\sqrt{2\pi}}\right)^2 \right), \nonumber
\end{equation}
where $Erfi$ is the imaginary error function \cite{Grad}
and therefore one sees that even for $c=1$ there are non-trivial
contributions to the anisotropy equation.

\item{(d)} We have considered the equation of motion where the
logarithm  terms were dominant, but
more physically for the non-relativistic case, we
can anticipate the scalar potential
to be the largest contributor to energy density. We
may again choose to set either
$F_1$ or $F_3$ fixed to be constant, which leads to the resulting
 expressions shown below;
\begin{eqnarray}
e^{-4\alpha - cF_1 F_2 \phi^2/M_p^2} &\sim& \frac{2 F_1 (1-c)}{c P_A^2} \hspace{0.8cm} (F_3) \nonumber \\
& \sim & \frac{(1-2c) e^{-2\alpha}}{P_A F_3 \sqrt{c}} \hspace{0.5cm} (F_1).
\end{eqnarray}
The first solution is constant for constant $F_3$ and clearly
vanishes for $c=1$. The second solution is for constant $F_1$, but
clearly decreases with time. Moreover the solution does not vanish
for $c=1$, rather it vanishes for $c=1/2$ - indicating that
anisotropies will be increasing in this case.

\end{description}

\subsubsection{\bf $AdS_5$ Solution}

As an example of a particular solution, let us consider the case of a pure $AdS_5$
embedding, generated by $N$ coincident $D3$-branes at large $N$. The background
functions must now satisfy the following expressions $F_1  =
\phi^4/\lambda^2, F_3 = 1/\sqrt{F_1}$ and $F_2$ is constant. Since the background functions
are explicitly known, the gauge coupling is fixed - in fact it is unity which leads to a
trivial result.
The solutions depend explicitly
upon the particular brane being embedded into the theory.

We focus initially on non-BPS configurations where $q \ne 1$. In
this case, we must ensure that $\lambda >> 1$ for analytic
solutions. We will again assume that the scalar potential is
quadratic. With the gauge field set to zero,  we find the equation
of motion at leading order in $\lambda$ becomes
\begin{equation}\label{eq:ads_eom}
\phi'(\alpha) \sim -\frac{2 M_p^2 \lambda^2}{F_2 \phi^5} + \ldots,
\end{equation}
where the higher order terms are sub-leading in $\lambda^2$. The above
equation admits the following solution
\begin{equation}\label{eq:chi}
\phi^6(\alpha) \sim \phi_0^6 - \frac{12 \alpha \lambda^2 M_p^2}{F_2}
\end{equation}
where we neglect the sub-leading terms in $\lambda$. This leads to a cancellation of the scalar field
mass in the equation of motion, therefore the solution does not appear to depend on it at leading order.
The corrections coming from the gauge field lead to the following master equation
\begin{equation}
\phi' (\alpha) \sim - \frac{2 M_p^2 \lambda^2}{F_2 \phi^5}
\left(1-\frac{\lambda^4 P_A^2 e^{-4\alpha}}{\phi^4} \right),
\end{equation}
which must be solved perturbatively. Let us define $\phi \equiv \chi
+ P_A^2 \xi + \ldots$ at leading order - where the solution for
$\chi$ is the one found above in (\ref{eq:chi}). The back-reacted
solution therefore has the leading order solution
\begin{equation}
\phi(\alpha) \sim \chi \left(1+ \frac{\lambda^2 e^{-F_2 \phi_0^6/(3
\lambda^2 M_p^2)}}{36 \chi^4} \left \lbrace \tilde{E} \left \lbrack
\frac{2}{3}, -\frac{F_2 \chi^6}{3 \lambda^2 M_p^2} \right \rbrack -
\frac{\phi_0^2}{\chi^2} \tilde{E} \left\lbrack \frac{2}{3},
-\frac{F_2 \phi_0^6}{3 \lambda^2 M_p^2}\right
\rbrack\right\rbrace\right) ,
\end{equation}
where we use the short-handed notation $\widetilde{E}[x,y] =
\rm{ExpIntegralE}[x,y]$ for simplicity \cite{Grad}. 
Note that the boundary conditions ensure that $\phi
\sim \phi_0$ at $\alpha = 0$. Since both $F_1$ and $F_3$ are defined
by the background geometry in this instance, we may find it hard to
satisfy our gauge density constraint (\ref{eq:R_DBI}). Indeed, one
can see that $\mathcal{R}$ only depends on the ratio
$e^{-4\alpha}/\phi^2$ which cannot be constant at this order of
approximation.

In the case of $q=1$, which is the BPS configuration (in the static
limit) we now need to assess whether we keep the $\dot{\phi}^2$
terms in the equation of motion. If we set them to zero, then we
recover the expression (\ref{eq:ads_eom}) as an exact result, not a
perturbative one. The back-reacted solution will also then follow
trivially. Instead let us keep the quadratic terms in the equation
of motion. We can then solve this equation in the absence of a gauge
field to obtain the expression
\begin{equation}
\dot{\phi} \sim - \frac{3 \dot{\alpha}\phi}{2} + \ldots,
\end{equation}
where we must ensure that the following condition is satisfied
\begin{equation}\label{eq:ads_constraint}
1 >> \frac{3 F_2 \phi^6}{16 \lambda^2 M_p^2}.
\end{equation}
This can be done if we also assume $\lambda^2 >> 1$, which is good for the supergravity approximation.
In this limit of the theory we find that dynamic solutions are exponentially decaying
\begin{equation}\label{eq:ads_constraint-a}
\phi \sim \phi_0 e^{-3\alpha/2}
\end{equation}
and if one includes the backreaction at leading order, the solution is
\begin{equation}
\phi(\alpha) \sim e^{-3\alpha/2}\left(\phi_0^4 \pm \lambda^2
P_A^2(1-e^{2\alpha}) \right)^{1/4},
\end{equation}
where we must again impose the constraint condition
(\ref{eq:ads_constraint}). Intuitively these results make sense,
because we are forced to consider a perturbative expansion in the
mass term, which effectively decouples the scalar potential from the
theory. The scalar field is driven purely by the potential generated
by the AdS background.

The energy density ratios in the
two cases of interest may be written
\begin{equation}
\mathcal{R}_1 \sim \frac{P_A^2 \lambda^3}{m^2 \phi^6}\left(
\frac{\phi_0}{\phi}\right)^{8/3}, \hspace{1cm} \mathcal{R}_2 \sim
\frac{P_A^2 \lambda^3}{m^2
\phi^6}\left(1-\frac{F_2(\phi_0^6-\phi^6)}{3\lambda^2 M_p^2}
\right),
\end{equation}
where $\mathcal{R}_1$ is the solution arising when we neglect
$\dot{\phi}^2$ terms in the equation of motion. Note that
$\mathcal{R}_1$ is initially very small, but increasing during
inflation indicating that the energy density of the gauge field is
highly suppressed. As $\phi$ continues to decrease, corresponding to
the limit where the probe branes are nearing the $D3$-branes, the
ratio rapidly starts to diverge
indicating that the gauge field contribution overwhelms the scalar
energy density and back-reaction dominates. The solution for
$\mathcal{R}_2$ also increases as the brane approaches the stack,
although one must be careful to ensure that the parameters satisfy
the constraint (\ref{eq:ads_constraint}). One notes that, when
compared to $\mathcal{R}_1$ with similar choices of parameters, that
$\mathcal{R}_2$ is significantly smaller in magnitude than
$\mathcal{R}_1$, and the gauge field domination occurs at later
times.

Regarding the anisotropies we see
that herein the equation of motion  in the BPS case $q=1$ (dropping
the $\dot{\phi}^2$ terms) can be written as follows;
\begin{equation}
\frac{\dot{\sigma}}{\dot{\alpha}} \sim \frac{4 P_A^2 \lambda^3}{3
m^4 \phi^8}\left( \frac{\phi_0}{\phi}\right)^{8/3},
\end{equation}
which are increasing rapidly in this instance due to the
exponentially decaying behaviour of the scalar field. In the case
where we neglect  the
$\dot{\phi}^2$ terms, the solution becomes
\begin{equation}
\frac{\dot{\alpha}}{\dot{\sigma}} \sim \frac{4 P_A^2 \lambda^2}{3
m^4 \phi^8}\left( 1 - \frac{F_2(\phi_0^6-\phi^6)}{3 \lambda^2
M_p^2}\right),
\end{equation}
at leading order in $\lambda$ for all $q$. The result is that the
anisotropies are increasing during inflation, recalling that the
parameter space is tightly constrained, and at a smaller level than
the previously considered solutions.

\subsubsection{\bf Tachyonic solution}

Let us consider the non-relativistic expansion for the tachyonic
solution in a non-BPS configuration, where $F_2$ is a constant,
$F_1$ is the tachyon potential and $q=0$. We will take the potential
to be of the form $F_1 \sim V_0/\cosh(\phi/L)$, where $L$ is an
unknown dimensionful parameter. We leave this arbitrary because it
is highly likely that the tachyon potential in curved space is
different from the one derived using BSFT (Boundary String Field Theory)
in flat space \footnote{See the recent paper \cite{Niarchos:2010ki} for
additional clarification}. Additionally  we will also set the scalar
potential to zero, ensuring that the inflaton dynamics is driven
only by open string condensation on the world-volume. The resulting
ratio constraint from (\ref{eq:R_DBI}) becomes
\begin{equation}
F_1^2 F_3^2 \sim Z^2 e^{-4\alpha c},
\end{equation}
where $Z$ is a dimensionful
parameter of the theory (cf. Appendix).

The background field equation with the above tachyonic potential results in the following
expression for the inflaton
\begin{equation}
\tanh \left( \frac{\phi}{2 L}\right) \sim \exp \left( \frac{\alpha
M_p^2}{2F_2 L^2 V_0}\right),
\end{equation}
where we have dropped the constants of integration. It is well known
that tachyonic inflation requires severe
fine-tuning, however the literature
has  have always assumed the validity of the flat space potential
in curved space. Indeed when warping is taken into consideration,
this fine tuning decreases allowing for potential inflationary
trajectories \cite{Raeymaekers:2004cu}. In our case the warping is
essentially embedded in the definition of $L$, which can lead to
inflation when it is sufficiently large since it suppresses the mass
of the inflaton.

The gauge coupling in this instance can be written
as follows
\begin{equation}
F_1 F_3^2 \sim \frac{Z^2}{V_0} e^{-4\alpha c} \cosh \left(2
   \rm{arctanh}\left(e^{\alpha r} \right) \right),
\end{equation}
where we defined $r\equiv
M_p^2/(2 F_2 L^2 V_0)$. Note that this coupling is non-positive, and
increasing towards zero from below. This is true even if the brane
was a ghost brane, with negative tension.
The backreacted equation of motion can be written as
\begin{equation}
\phi'(\alpha) \sim -\frac{M_p^2}{2 F_2} \left( 1+\frac{4x}{3}\right)
\frac{\partial}{\partial \phi} \ln F_1,
\end{equation}
which admits the following asymptotic solution
\begin{equation}
\tanh\left( \frac{\phi}{2 L}\right) \sim
\frac{1}{|\rm{coth}\left(\frac{\epsilon}{2L} \right)|} \exp
\left(\frac{M_p^2(P_A^2 (-1 +
e^{4\alpha(c-1)})+3\alpha(c-1)Z^2))}{6(c-1)F_2 L^2 V_0 Z^2}\right),
\end{equation}
where we have included the initial boundary condition on $\phi$ such that $\phi \to \epsilon <<L $ at $\alpha =0$.
The gauge coupling is, in general, a rather complicated function of the scale factor
\begin{equation}
F_1 F_3^2 \sim \frac{Z^2 e^{-4\alpha}}{V_0} \cosh \left(2
\rm{arctanh}\left( tanh\left( \frac{\epsilon}{2L}\right) \exp \left(
\frac{r}{3Z^2(c-1)}\left( P_A^2(-1+e^{4\alpha(c-1)})+3\alpha
Z^2(c-1))\right) \right)\right) \right),
\end{equation}
for arbitrary values of $c$; However,  for $c=1$ the solution
reduces to
\begin{equation}
F_1 F_3^2 \sim \frac{Z^2 e^{-4\alpha}}{V_0} \cosh \left(2
\rm{arctanh}\left( tanh\left( \frac{\epsilon}{2L}\right)\exp
\left(\frac{r\alpha}{3Z^2}[4P_A^2 + 3 Z^2] \right)\right) \right),
\end{equation}
where $r$ is the same quantity defined earlier. Because of the
perturbative correction in $P_A$ the gauge coupling  for $c=1$ can
be a positive, decreasing function (as a function of the expansion).
However due to the algebraic structure of the solution, there is a
critical value of the scale factor
\begin{equation}
\alpha_c = \frac{3Z^2}{r (4 P_A^2 + 3Z^2)} \ln \left( \rm{coth} \left(\frac{\epsilon}{2L}\right)\right)
\end{equation}
at which the solution is singular, therefore we can only trust the
solution for $\alpha < \alpha_c$. When the background is uniquely
specified in this case, we see that decreasing solutions require
$r<<1$. When we initially fix $r$, we then see that a decreasing
gauge function requires $Z^2 << V_0$.

\subsection{Relativistic limit}
\label{rel-lim}

\indent

The relativistic limit occurs when $\gamma >> 1$, corresponding to
$\dot{\phi}^2 \sim 1/F_2$, and the scalar equation of motion takes
the following form
\begin{eqnarray}
 0 & \sim& F_2 \gamma \dot{\phi} \left(1-\frac{x}{2} \right) \left(3 \dot{\alpha} + \dot{\phi} \frac{\partial}{\partial \phi} \ln(F_1) + \frac{\dot{\phi}(1+\gamma^2 F_2 \dot{\phi}^2)}{2}\frac{\partial}{\partial \phi}\ln(F_2) \right)  \\
&+& F_2 \gamma \dot{\phi}x \left(2\dot{\alpha} + \dot{\phi}
\frac{\partial}{\partial \phi} \ln(F_1 F_3) \right) +
\frac{1}{\gamma} \left(1-\frac{x}{2} \right)
\frac{\partial}{\partial \phi}\ln(F_1) -\frac{x}{\gamma}
\frac{\partial}{\partial \phi} \ln(F_1 F_3) + \frac{V'-qF_1'}{F_1}.
\nonumber
\end{eqnarray}
Since $\gamma$ is controlled by $F_2$, and we know the relation
between this function and the inflaton,  we will treat this as an
unknown variable allowing us to eliminate the velocity from the
problem. The equation of motion can then be solved once we specify
the potential and the function $F_1$. Unlike the non-relativistic
limit, there is no `trivial' simplification that one can consider.
The best approach turns out to be fixing the potential to be
quadratic, as before, and assuming that $F_1$ satisfies power law
behaviour, $F_1 \sim L \phi^p$ where $L$ is herein a constant with
mass dimension $(4-p)$.

In general, our assumption that $\gamma >> 1$ allows us to solve the
system explicitly once we specify $F_2$. To illustrate, let us
assume that $F_2 \sim \overline{W} \phi^p$. We can then integrate
the equation of motion directly to obtain
\begin{eqnarray}
\phi(t) &\sim& \frac{m}{M_p} \frac{1}{\sqrt{6}(p+4)} \left(\sqrt{W}
\phi_0^{(2+p)/2} \left(\phi_0 - \phi \right) +t \phi(2 + p) \right),
\hspace{1cm} p \ne -4 \\ &\sim&
\frac{\sqrt{T}\phi_0}{\sqrt{T}+t\phi_0}, \hspace{7.8cm} p=-4
\end{eqnarray}
where $\overline{W}, T$ are constants.
We can then investigate the
following cases:

\begin{description}

\item{(a)} Let us  assume that $V
>> F_1 q$ as a constraint on the Hubble equation. This is the limit
in which (canonical) DBI-inflation can occur \cite{Silverstein:2003hf, Alishahiha:2004eh}. Physically this corresponds
to the inflaton being strongly damped by the Hubble factor, so that
even though it is relativistic in velocity, it does not travel a
large distance in field space. Knowing the scalar field solution, we
can then use the Hubble equation to determine the scale factor
\begin{eqnarray}
\alpha(\phi) &\sim& \frac{m}{M_p}\frac{1}{\sqrt{6}(4+p)}
\left(\sqrt{T}\phi_0^{(2+p)/2}(\phi_0-\phi) + t\phi(2+p) \right),
\hspace{1cm}p \ne -4 \\ &\sim& \frac{m}{M_p} \sqrt{\frac{T}{6}}\ln
\left( \frac{\phi_0}{\phi}\right), \hspace{6.4cm} p=-4
\end{eqnarray}
where we have assumed that $\phi$ is non-negative for simplicity - and imposed boundary conditions on the scale factor so that it vanishes
at the start of inflation.

One could include the gauge field corrections to the above results,
by employing the relevant term in the Hubble parameter. However it
will turn out to be much simpler to use the scalar field equation to
solve for $F_2$, and then compute the back-reaction on this
variable. We will now consider this strategy where $F_2$ is unknown
and $F_1$ has power law behaviour $F_1 \sim L \phi^p$. The general
equation is quite difficult to
solve analytically once one includes the back-reaction. Therefore we
illustrate the results for the case $p=2$, which has a leading order
expansion of the form
\begin{equation}
F_2 \sim \frac{2 M_p^2 \gamma^4(\nu_2-2)^2}{3 m^2
\phi^4}\left(1+\frac{2 P_A^2}{3 L \phi^2} \frac{(12+(\nu_2-2)(4c
\pm1))}{(4\gamma^3(c-1)(\nu_2+2)-3\nu_2)}\left(\frac{\phi_0}{\phi}
\right)^{4\gamma^3(c-1)(\nu_2-2)/3} \right),
\end{equation}
where the $\pm$ sign arises from the particular choice of sign in
the Hubble equation, and $\nu_2 \equiv \frac{m^2}{\gamma^3 L}$. One
notes that the leading order dependence on the inflaton is $1/
\phi^4$ and therefore one expects that the scale factor should
initially run logarithmically (at leading order)
\begin{equation}
\alpha \sim \frac{\gamma^2|\nu_2 -2|}{3} \ln
\left(\frac{\phi_0}{\phi} \right),
\end{equation}
which can rapidly become large due to the overall pre-factor of
$\gamma^2$, and the constraint that $V >> F_1 \gamma$ always ensures
that $\nu_2 >> 2$.

We can immediately ask what happens to the anisotropies in this
limit. Indeed, in the non-relativistic limit we discovered that they
increased during the inflationary epoch because of the back-reaction.
A quick calculation in the
relativistic limit implies
\begin{equation}
\frac{\dot{\sigma}}{\dot{\alpha}} \sim \frac{\gamma P_A^2}{3 m^2
\phi^2}\left(\frac{\phi_0}{\phi} \right)^{4\gamma^3(\nu_2-2)(c-1)/3}
,
\end{equation}
where $c$ is the parameter arising from the gauge field condition
$F_1 F_3^2 \sim e^{-4\alpha c}$. One can therefore see that anisotropies increase quite
rapidly in this instance - significantly faster than in the non-relativistic limit due to
the dependence of $\gamma$ in the exponent.


\item{(b)} The converse limit, where $F_1 \gamma >> V$ is also interesting,
since - like the tachyonic theory in the non-relativistic limit -
the gauge field constraint in (\ref{eq:R_DBI}) imposes the condition
that $F_1^2 F_3^2 \sim Z^2 e^{-4\alpha c}$. We will again assume
that $F_1$ is power law to simplify the equations of motion. The
isotropic theory admits the following solutions
\begin{eqnarray}
F_2 &\sim& \frac{M_p^2(\gamma^2(2+p)-2p)^2}{12 \gamma L \phi^{2+p}}, \\
\alpha &\sim&
\frac{(\gamma^2(2+p)-2p)}{6}\ln\left(\frac{\phi_0}{\phi} \right),
\end{eqnarray}
which are significantly different from the potential dominated
regime. In particular we see that the scale factor is logarithmic
for all values of $p$. This solution actually has many similarities
to the non-relativistic tachyon theory, since we can again write the
gauge coupling function as follows
\begin{equation}
F_1 F_3^2 \sim e^{-Q_1\alpha}, \hspace{1cm} Q_1 \equiv 4c +
\frac{6}{\gamma^2(2+p)-2p}. \label{F1F3-b-r}
\end{equation}
Including the gauge field back-reaction, we then obtain the leading
order solution
\begin{equation}
F_2 \sim \frac{M_p^2(\gamma^2 (2+p)-2p)^2}{12 \gamma L
\phi^{2+p}}\left(1 + \frac{P_A^2}{Z^2}\left(
\frac{\phi_0}{\phi}\right)^{2(c-1)(\gamma^2(2+p)-2p)/3}
\frac{(8c+3\pm1)}{(4 \gamma^2(c-1) \mp 3)} +\ldots \right),
\end{equation}
where the correction term will clearly be constant when $c=1$. The
choice of sign again arises from the Hubble expression. Upon
integration we then find the following solution for the scalar field
\begin{equation}
\frac{\phi(\alpha)}{\phi_0} \sim e^{-6 \alpha/\mathcal{N}}\left(1+
\frac{3 P_A^2}{4
Z^2(c-1)\mathcal{N}}\frac{8c+3\pm1}{4\gamma^2(c-1)\mp3} \right),
\end{equation}
which simplifies considerably when $c=1$ to become
\begin{equation}
\frac{\phi(\alpha)}{\phi_0} \sim
e^{-6\alpha/\mathsf{N}}\left(1+\frac{3 P_A^2
\alpha}{\mathsf{N}Z^2}\frac{12\pm1}{\mp3} \right),
\end{equation}
where we have herein used the notation $\mathsf{N} \equiv
\gamma^2(2+p)-2p$. This latter expression can be inverted to obtain
the scale factor in terms of the Lambert W function \cite{Grad}, which admits the following
expansion;
\begin{equation}
\alpha \sim \frac{\mathsf{N} Z^2}{3 P_A^2
\delta}\left(\frac{\phi}{\phi_0}-1 \right), \hspace{1cm}\delta
\equiv  \frac{12\pm1}{\mp3},
\end{equation}
at leading order in $P_A^2$. The solution for the scale factor
therefore appears to be non-perturbatively corrected by the presence
of the gauge field. The gauge coupling then takes the schematic form
\begin{equation}
F_1 F_3^2 \sim \alpha^{-p} e^{-4\alpha c},
\end{equation}
which exhibits a minimum if $\alpha_c = -p/(4c)$ where $p, c$ have
opposite sign and $c <0$ - however since we require $c>0$ for
physical solutions, this is most likely not a physical result.
However this point is a maximum when $p < 0$ with $c>0$, thus there
are solutions where the gauge coupling initially grows in strength,
before reaching its maximal value at which point it starts to
decrease. This unusual behaviour arises from the analytic structure
of the DBI action, and therefore has no natural analog in terms of
canonical scalar field models.

\item{(c)} Let us briefly discuss the tachyonic solution in the relativistic limit.
The isotropic equation of motion yields the following
solution for the scale factor valid at large $\gamma$
\begin{equation}
\alpha = \frac{1}{3} \ln \left( \cosh
\left(\frac{\phi}{L}\right)\right). \label{c-r}
\end{equation}
Including the leading order
back-reaction we then find the following solution
\begin{equation}
\cosh\left( \frac{\phi}{L}\right) \sim \left( \sqrt{\frac{Z}{P_A}}
W\left\lbrack \frac{P_A^2}{Z^2}
e^{12\alpha(c-1)}\right\rbrack^{1/4}\right)^{1/(c-1)},
\end{equation}
in terms of the Lambert W function, and valid for all $c \ne 1$.
This modifies the gauge coupling function resulting in the
expression
\begin{equation}
F_1 F_3^2 \sim \frac{Z^2}{V_0}e^{-4\alpha c}\left(
\sqrt{\frac{Z}{P_A}} W\left\lbrack \frac{P_A^2}{Z^2}
e^{12\alpha(c-1)}\right\rbrack^{1/4}\right)^{1/(c-1)},
\end{equation}
which is not necessarily a decreasing function. For $c < 3/4$ we see
that this function actually increases with scale factor, whilst for
$c>3/4$ the function is monotonically decreasing, with an amplitude
set by the ratio $Z^2/V_0$. For sufficiently small electromagnetic
energy density we can expand the Lambert function and we then find
that
\begin{equation}
F_1 F_3^2 \sim \frac{Z^2}{V_0}(-\eta \gamma H)^{3+c},
\end{equation}
which corresponds to a spectral index of $n=7+c$ for the density
ratio, which should be contrasted with the tree-level result which
yields $n=1+4c$. Therefore we see that even a tiny electromagnetic
field in the tachyon case leads to additional dependence on the
expansion parameters, and a significantly larger magnetic field for
$c< 2$.

\end{description}

\section{Discussion and Outlook}
\label{DandO}

In this paper we have investigated
a class of corrections to inflationary solutions arising from the
introduction of an electromagnetic
field in a non-linear context. We
first considered the pure Einstein-Born-Infeld theory, and found
that, in order to have satisfactory inflation,  we needed to promote
the BI coupling to have scalar
field dependence. For power law solutions we then used the WMAP data
set to restrict the functional form of the
coupling, subsequently retrieving a
range of parameters inducing models observationally consistent.

Following this, we employed a
generalised approach to $D$-brane inflation
where space-time
backgrounds \footnote{Such backgrounds are generated by gluing warped
throats onto a Calabi-Yau three-fold.} 
were \emph{not} necessarily
asymptotically AdS (cf. \cite{Bamba:2008my} for an analysis
including an asymptotically AdS setting.)
This allowed for more richer classes
of couplings than in pure Einstein-Born-Infeld settings. In all
cases, the presence of the gauge field was subleading in the
corresponding inflationary trajectory. Moreover, anisotropies
 tend generically to increase with
time; The anisotropies were treated perturbatively in this paper,
but a more detailed analysis in such models would be welcome. The
one case where this was not occurring  was when the logarithmic
terms dominate the non-relativistic equation of motion: The
anisotropies decrease with time, provided that the scalar potential
satisfies a (stringent) non-trivial bound. 

Concerning the analysis above
summarized, it is of relevance point out that the scalar potential
was assumed to be of a simple power law form. Whilst this is
convenient for comparing solutions to the canonical
(non-relativistic) limit, it is not
necessarily easy to obtain such solutions within a string theory
context: Although quadratic potentials are common within
supergravity theories, stringy instantons typically give rise to
exponential potentials; Therefore some of the results obtained in
this paper will not be valid in a more complete string theory
embedding.

Finally, regarding the evolution of
magnetic seed fields in a DBI context \footnote{The magnetic field on
the $D3$-brane can be seen to be generated by exciting the open
string degrees of freedom. Since the $F$ and $D$ strings are S-dual,
the most general configuration would be a mixture of both string
solutions. Since the increase in world volume flux tends to increase
the mass of the moving brane, it is likely that the brane will
'sink' lower in the throat therefore making inflation harder to
occur. However there may well be a set of backrounds where this does
not occur. Turning on such flux, however, could be useful for
transferring inflationary energy to the standard model sector
(should they be localised on seperate branes), because the open
strings on the inflationary $D3$-brane could attach themselves to
the SM branes at the end of inflation. Such a process would
correspond to the direct transmutation of energy from inflaton
sector to the standard model degrees of freedom.}, a brief analysis
was provided in Appendix A. The essential feature is the new
couplings that the DBI configuration induces. It was seen that the
electromagnetic coupling depends on two parameters. We chose to fix
one of these parameters by demanding that the energy density ratio
was constant. Two classes of solution could then be identified. The
first was a generalised version of the canonical coupling, given by
$e^{-4 \alpha c}$, which led to scale invariance when $c=3/2$. The
second class of solutions were specific to the model under
consideration. Notably the tachyon (relativistic and
non-relativistic limits) and the ultra-relativistic limit of the
$D3$-brane theory. In such models the scale invariance arose for
values of $c < 1$, as demonstrated explicitly in the case of the
tachyon which has a coupling of the form $e^{\alpha (3+c)}$. For
special classes of DBI models in the relativistic limit we found
that the coupling behaved like $\alpha^{-p} e^{-4\alpha c}$ where
$p$ arose from the assumption of power law dependence. Overall, we
found that the obtained magnetic field spectrum is typically larger
than the observed bound unless it was created primordially (with
subsequent amplification via the dynamo mechanism).

As a last note, allow us to indicate
that  a maximal bound on the strength of the gauge field could be
generically established. When saturated, implying that a non-zero
field is generated at sufficiently small scale,  a residual
cosmological constant emerges - independent of the form of the
scalar potential. It is not, therefore, inconceivable that a theory
could eventually contribute to resolve the the dark energy problem
(and possibly the issue of primordial magnetogenesis).

\indent

\appendix

\section{Magnetic field generation} \label{mag gen}

\indent

We have explicitly considered a non-zero electric field component as
a solution to the field equations (\ref{3.12})-(\ref{3.14}) in the
main body of the paper. For the DBI theory of $D3$-branes this
corresponds to turning on $F$-string flux on the world-volume, and
is therefore rather natural. The magnetic field in this case
corresponds to the $D1$-string flux. The $D$-brane theory exhibits
S-duality invariance, which encompasses the electro-magnetic duality
of ordinary Maxwell theory.
Consequently,  one can deduce the
levels of the magnetic field from the electric field, and
vice-versa.

Let us therefore consider the manifestation of such fields during
inflation. In general this is a highly technical problem,
hence we must consider simplifying
the background to make the analysis tractable. This is usually done
by initially consideration of the flat FRW geometry (cf. section 1),
and moving to a conformal gauge
such that $\tau \equiv  \int e^{-\alpha} dt$. The gauge field can
then be expanded into annihilation/creation operators and mode
functions ${\mathsf{A}}(\tau, k)$ satisfying the Fourier space
equation
\begin{equation}
\frac{\partial^2 {\mathsf{A}}}{\partial \tau^2} +
\frac{\partial}{\partial \tau}\ln(F_1 F_3^2 \gamma) \frac{\partial
{\mathsf{A}}}{\partial \tau} + \frac{k^2}{\gamma^2} {\mathsf{A}} =
0,
\end{equation}
where $k$ is the comoving wavenumber. If we now introduce another
change of variables $\eta \equiv  \int d\tau \gamma^{-1}$, then the
above equation becomes the more familiar wave equation
\begin{equation}
\frac{\partial^2 {\mathsf{A}(\eta, k)}}{\partial \eta^2} + 2
\frac{\partial }{ \partial \eta}\ln(F_1 F_3^2) \frac{1}{2}
\frac{\partial {\mathsf{A}(\eta, k)}}{\partial \eta} + k^2
{\mathsf{A}(\eta, k)} = 0
\end{equation}
and we see that $\eta = - e^{-\alpha}/(\gamma H)$ for constant
$\gamma H$. This definition allows us to divide the mode functions
into sub and super horizon modes, where the former occurs for $k>>
\gamma H e^{\alpha}$ and the latter for modes satisfying $k <<
\gamma H e^{\alpha}$, for some characteristic time scale $\eta_k
\sim -1/k$. Making the following identification
\begin{equation}
\frac{Y'}{Y} \equiv \frac{1}{2} \frac{\partial \ln (F_1
F_3^2)}{\partial \eta},
\end{equation}
using primes herein to denote derivatives with respect to $\eta$, we
obtain the following Schr\"{o}dinger type expression through the
change of variables $v(\eta, k) = Y(\eta) {\mathsf{A}}(\eta, k)$:
\begin{eqnarray}
0 &=& v'' + v \left(k^2 - \frac{Y''}{Y} \right)  \\
&=& v'' + v \left( k^2 - \frac{F_3''}{F_3}- \frac{F_1''}{2 F_1} -
\frac{F_3'}{F_3}\frac{\partial}{\partial \eta} \ln(F_1^2
F_3)\right).
\end{eqnarray}
Sub-Hubble modes admit a solution given by the WKB approximation,
written in terms of the gauge field mode expansion ${\mathsf{A}}$
\begin{equation}
{\mathsf{A}}_{\rm in} \sim \frac{e^{-ik\eta}}{\sqrt{2 k F_1 F_2^3}},
\end{equation}
whilst the super-Hubble modes admit a solution of the form;
\begin{equation}
{\mathsf{A}}_{\rm out} \sim C_1(k) +C_2(k) \int \frac{d\eta}{F_1
F_3^2},
\end{equation}
where the $C_i$ are constants of integration which can be determined
by matching the in and out modes (and their derivatives) at horizon
crossing $H e^{\alpha} = k$. Note the above expression is valid at
leading order in $k$. Neglecting the subsequent decay mode for in
${\mathsf{A}}_{\rm out}$ one can then write the following general
solution for the mode expansion - following the arguments presented
in \cite{Bamba:2008my}
\begin{equation}
|{\mathsf{A}}(\eta, k)|^2 \sim \frac{1}{2 k (F_1
F_3^2)_{*}}\left|1-\left(i + \frac{(F_1 F_3^2)_{*}'}{2 k (F_1
F_3^2)_{*}} \right)k\int_{\eta_k}^{\eta_R} \frac{(F_1 F_3^2)_{*}
d\eta'}{(F_1F_3^2)}\right|^2,
\end{equation}
where the asterisk subscript denotes that the quantity is evaluated
at horizon crossing. Assuming that $\gamma, H$ are both constant
during inflation, which is a good approximation, then we can use the
following identity
\begin{equation}
d\eta = \frac{e^{-\alpha} d\alpha}{\gamma H}
\end{equation}
to simplify the above expression. One can calculate the vector field
correlation function to obtain the electric field power spectrum,
and using the inverse Maxwell relation $B_i =
\epsilon_{ijk}F^{jk}/2$ we then see that we can write the (proper)
magnetic field power spectra in the following manner
\begin{equation}
|B|^2 \sim 2 k^2 e^{-4 \alpha} |A|^2,
\end{equation}
which has the correct behaviour in the radiation dominated epoch so
that $B \sim e^{-2 \alpha}$. The derivation of this expression
relies on the fact that we assume instantaneous reheating, and that
the conductivity becomes much larger than $H$ immediately after
inflation. Introducing the density parameter $\Omega(\eta ,k) =
\rho_B/\rho_{\gamma}$ where $\rho_{\gamma}$ is the energy density of
radiation, and $\rho_B$ is the gauge field energy density (per unit
logarithm), we have the definition;
\begin{equation}
\Omega(\eta, k) = \frac{15}{2 \pi^4 N}\left( \frac{k
e^{-\alpha_R}}{T_R}\right)^4 \frac{(F_1 F_3^2)}{(F_1
F_3^2)_*}\left|1-\left(i + \frac{(F_1 F_3^2)_{*}'}{2 k (F_1
F_3^2)_{*}} \right)k\int_{\eta_k}^{\eta_R} \frac{(F_1 F_3^2)_{*}
d\eta'}{(F_1F_3^2)}\right|^2 ,
\end{equation}
where $N$ is the number of massless degrees of freedom at reheating,
and $T_R$ is the reheating temperature
(a subscript $R$ denotes the
quantity evaluated at the reheating epoch). Assuming that the gauge
coupling exhibits power law behaviour such that
\begin{equation}
F_1 F_3^2 \sim {\rm const} \times \left(
\frac{\eta}{\eta_0}\right)^{-p},
\end{equation}
we can then see that
\begin{equation}\label{eq:Omega}
\Omega(\eta, k) \sim N \left( \frac{T_R}{M_p}\right)^4 \left(
\frac{k e^{-\alpha_R}}{H_R}\right)^{4-p} F_1 F_3^2 \gamma_R^p (F_1
F_3^2)_R^{-1},
\end{equation}
indicating that the spectral index is given by $4-p$. The precise
value of $p$ can then be determined once the precise theory is
specified.

At scales of order $L=2\pi/k$ [Mpc], it was subsequently
derived in \cite{Bamba:2008my}  the
following bound for the present magnetic field assuming a reheat
temperature of the order $10^{15} GeV, N=100$
\begin{equation}\label{eq:B_field_today}
|B|_0 \sim 10^{11p-57}
\sqrt{W(p)\frac{F(0)}{F(\eta_R)}}\gamma_R^{p/2}\left(\frac{L}{\rm{Mpc}}\right)^{(p-4)/2},
\end{equation}
where we have defined the gauge function as $F_1 F_3^2 = F(\eta)$
and $W(p)$ is a $p$-dependent function that diverges as $p \to -1$
and asymptotes to $W \sim 0.25$ as $p \to \pm \infty$. Note that we
must have $|B|_0 \ge 10^{-9} GeV$ to be consistent with observed
field strengths up to $1$-Mpc without envoking a dynamo mechanism.
Clearly this is a very tight bound for all models to satisfy. The
expression for the field at decoupling ($\alpha_{dec} \sim
-3\ln(10)$) is essentially the same. The only difference is that we
have a shifted overall exponent $10^{11p-51}$ and must of course
replace $F(0)$ by $F(\rm{dec})$, with
\begin{equation}\label{eq:B_field_decoupling}
|B|_{\rm dec} \sim 10^{11p-51}
\sqrt{W(p)\frac{F(\rm{dec})}{F(\eta_R)}}\gamma_R^{p/2}\left(\frac{L}{\rm{Mpc}}\right)^{(p-4)/2},
\end{equation}
which must now satisfy the bound $|B|_{\rm dec} \ge 10^{-23}G$ for
cosmic seed fields. This is a less stringent bound because in this
instance we can assume there is a dynamo mechanism which serves to
amplify the field at late times.


In terms of the gauge field coupling function
$F_1 F_3^2 $, we can determine two
immediate situations of interest (cf. eq. (\ref{eq:R_DBI})):
\begin{itemize}
\item Firstly,  there is the case  $F_1 F_3^2 \sim
e^{-4\alpha c}$ for both relativistic and non-relativistic cases,
where $c$ is a constant. This occurs when  the scalar potential
dominates the total energy density of the model. This is similar to
the condition established in the BI section and it can be shown that
the mode function after horizon crossing is proportional to
$e^{(4c-1)\alpha}$ - indicating that we must consider positive $c$,
because negative $c$ leads to a decreasing gauge field component,
which we naturally expect to have a smaller contribution to the
total energy density. Inserting this expression into the mode
expansion of the gauge field, and obtaining the result for
the proper magnetic field, one sees that (at the end of inflation)
\begin{equation}
|B|^2 \sim H_I^4e^{2 \alpha_f(2c -3)},
\end{equation}
where $H_I$ is the value of the Hubble parameter at horizon
crossing. For a flat spectrum we therefore recover the constraint
that $c=3/2$. Indeed one can show that this result implies that the
energy density of the electromagnetic sector rapidly becomes
comparable to that of the inflaton. For the marginal case where
$c=1$, we can explicitly solve the gauge field equation of motion.
Matching the solution to the vacuum in the sub-horizon limit, we can
identify the integration constants, and write the solution in terms
of Hankel functions
\begin{equation}
{\mathsf{A}}(k, \eta) \sim
e^{-ic\pi}\sqrt{\frac{\eta^{(1-4c)/2}}{2(-\gamma H)^{4c}}}
H^{(2)}\left(\frac{4c-1}{2},k\eta \right)
\end{equation}
which is plotted in Figure 2. In
terms of magnetic field spectral index and written as $n \equiv 4
-p$, we see that this immediately implies $n = 4(1+c)$. Since this
is positive definite for $c > 1$ we find that the spectral index is
highly blue-tilted and damped for larger values of $c$. Note that
$c=3/2$ yields an approximately flat spectrum as anticipated.
\begin{figure}
\begin{center}
\includegraphics[width=0.7\textwidth]{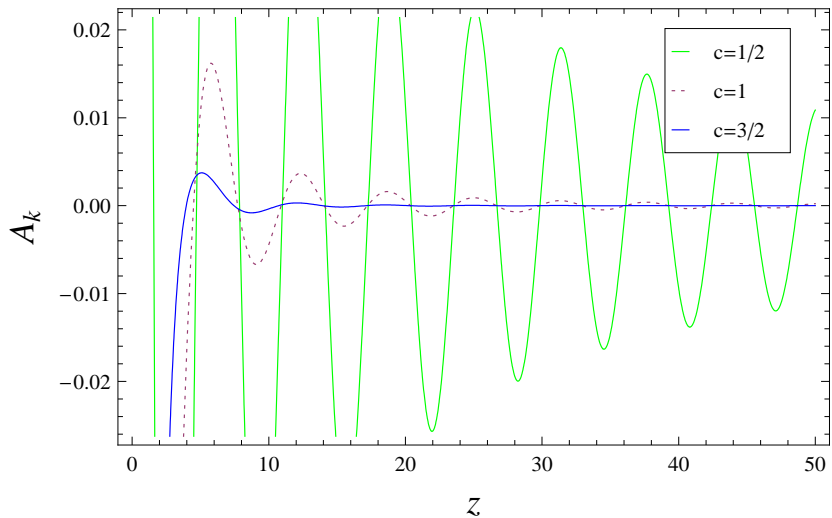}
\caption{Figure showing $Re(A) k^{(1-4c)/2}$ as a function of $z (=
-k\eta)$ for different $c$. We have also absorbed a factor of
$(\gamma H)^{4c}$ into the gauge field for simplicity. The level of
damping increases with $c$.}
\end{center}
\end{figure}

\item The second case of interest occurs when $F_1(\gamma - q) >> V$, which could arise
in the ultra-relativistic regime where $ F_1 \gamma$ is very large.
The constraint equation (\ref{eq:R_DBI}) then implies that $F_1^2
F_3^2 \sim Z^2 e^{-4\alpha c_1}$, where $c_1$ is a dimensionless
constant and $Z$ is a dimensionful constant. Since the gauge
coupling depends only on $F_1 F_3^2$, we must specify one of these
two functions completely, before being able to determine the
magnetic field spectrum. This \emph{new} limit is only possible due
to the non-linear nature of the DBI action, but appears to depend
explicitly on one of the background functions - unlike the previous
case. Interestingly this is precisely the combination of parameters
that determines the scale of the anisotropies in
(\ref{eq:anisotropies-a}). Our emphasis
has been  on determining $F_3$ for
a given value of $F_1$ unless otherwise stated.
Furthermore:

\begin{itemize}

\item Empolying (\ref{F1F3-b-r}) induces solutions of the form
\begin{equation}
{\mathsf{A}}(k,\eta) \sim \eta^{(1-Q_1)/2}\left(A_1 {\rm J} \left(
\frac{Q_1-1}{2}, k\eta \right) + A_2 {\rm Y} \left( \frac{Q_1-1}{2},
k\eta \right) \right),
\end{equation}
where $Y$ and $J$ are Bessel
functions \cite{Grad}. The assumption of large $\gamma$ ensures
that $Q_1$ tends to $4c$ from above for all values of $p>0$. In
order to obtain smaller values of $Q_1$ we must ensure that $p$ is
negative - in particular one can see that the $\gamma$ dependence
drops out for $p=-2$ which is plotted in Figure 3, although note
that $Q_1 \to -4$ asymptotically. The solution simplifies in the
case of $Q_1 = 1$, which occurs when $p$ takes the critical value
$p_c$;
\begin{equation}
p_c \sim \frac{\gamma^2(2-8c)-6}{(\gamma^2-2)(4c-1)},
\end{equation}
which asymptotes to $p_c = -2$ for large enough $\gamma$ independent
of the value of $c$. Therefore, if $Q_1$ is constrained to be unity,
then we see that $p \to -2$ is an attractor solution. Normalising
the solution using the sub-Horizon modes, we can extract the
coefficients arising from integration and write the solution:
\begin{equation}
{\mathsf{A}}_k \sim e^{-i\pi Q_1/4}
\sqrt{\frac{\eta^{1-Q_1}}{2(-\gamma
H)^{Q_1}}}H^{(2)}\left(\frac{Q_1-1}{2}, k\eta \right),
\end{equation}
where $H^{(2)}$ is the Hankel function of the second kind, which is
illustrated in Figure (3). Note that the attractor solution $p=-2$
ensures that the gauge field is strongly damped. Other values of $p$
tend to lie on the same curves, indicating that the solution is
insensitive to the precise form of the power law.
\begin{figure}[htp]
\begin{center}
\includegraphics[width=0.55\textwidth]{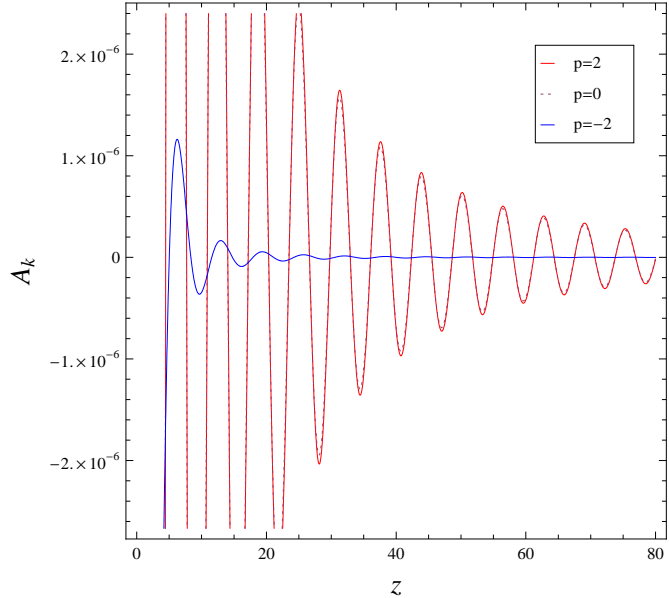}
\caption{The real part of the expression $A \sqrt{k^{1-Q}}$ plotted
as a function of $z (= -k\eta)$ for $c=1$. We assumed that $H^2 \sim
\gamma/3$ and $\gamma \sim 10$ using Planckian units.}
\end{center}
\end{figure}

\item From (\ref{c-r}),
 we can use it to solve the gauge field energy density constraint
to obtain the electromagnetic field as before. The solution is again
a Hankel function, but if we solve for the magnetic field we obtain
the following scaling
\begin{equation}
|B_k|^2 \propto H_I^4 H_I^{(3-4c)},
\end{equation}
which indicates that $c=3/4$ yields a flat spectrum at the end of
inflation.

\end{itemize}
\end{itemize}

There is also an interesting fixed point solution where we can find
$F_1 F_3^2 = $ constant. In this case we find that a magnetic field
emerges during the inflationary phase due to a logarithmic term in
the mode expansion of $\mathsf{A}$. Thus the field is driven to be
initially large, but decreases rapidly as the universe expands. The
gauge field then vanishes identically at the end of inflation -
therefore is unable to act as a seed-field to generate the observed
magnetic field today.

\section*{Acknowledgments}
\label{A}

PVM acknowledges the support of the grant CERN/FP/109351/2009. JW is supported
in part by NSERC of Canada.

\indent


\end{document}